\documentclass[a4paper,11pt]{article}
\pdfoutput=1 

\usepackage{jheppub} 

\usepackage[T1]{fontenc} 
\usepackage{xcolor}
\usepackage{url}
\usepackage{dsfont}
\usepackage{soul} 

\newcommand{\mgamc}{\textsc{MadGraph5\_aMC@NLO}}
\newcommand{\pt}{\ensuremath{p_{\mathrm{T}}}}
\newcommand{\fszero}{\ensuremath{f_{\mathrm{S}0}}}
\newcommand{\fsone}{\ensuremath{f_{\mathrm{S}1}}}
\newcommand{\fstwo}{\ensuremath{f_{\mathrm{S}2}}}
\newcommand{\fmzero}{\ensuremath{f_{\mathrm{M}0}}}
\newcommand{\fmone}{\ensuremath{f_{\mathrm{M}1}}}
\newcommand{\fmtwo}{\ensuremath{f_{\mathrm{M}2}}}
\newcommand{\fmthree}{\ensuremath{f_{\mathrm{M}3}}}
\newcommand{\fmfour}{\ensuremath{f_{\mathrm{M}4}}}
\newcommand{\fmfive}{\ensuremath{f_{\mathrm{M}5}}}
\newcommand{\fmseven}{\ensuremath{f_{\mathrm{M}7}}}
\newcommand{\kappav}{\ensuremath{\kappa_{2\mathrm{V}}}}
\newcommand{\mysigma}{\ensuremath{\overline{\sigma}}}
\newcommand{\mvv}{m_{VV}}
\newcommand{\ggzzh}{ggZZH}

\title{\boldmath Sensitivity to New Physics in final states with multiple gauge and Higgs bosons}

\author[a]{A. Cappati,}
\author[b,c]{R. Covarelli,}
\author[b,c]{P. Torrielli,}
\author[d,e]{M. Zaro.}

\affiliation[a]{Laboratoire Leprince-Ringuet, CNRS/IN2P3, Ec\'ole Polytechnique, \\ 
Institut Polytechnique de Paris, Av.\,Chasles, 91120 Palaiseau, France}
\affiliation[b]{Dipartimento di Fisica, Universit\`a di Torino,  \\
Via P. Giuria 1, 10125 Turin, Italy}
\affiliation[c]{INFN, Sezione di Torino, \\
Via P. Giuria 1, 10125 Turin, Italy}
\affiliation[d]{TIF Lab, Universit\`a di Milano, \\
Via Celoria 16, 20133 Milan, Italy}
\affiliation[e]{INFN, Sezione di Milano, \\
Via Celoria 16, 20133 Milan, Italy}

\emailAdd{alessandra.cappati@cern.ch}
\emailAdd{roberto.covarelli@cern.ch}
\emailAdd{paolo.torrielli@to.infn.it}
\emailAdd{marco.zaro@mi.infn.it}

\abstract{We analyse the sensitivity to beyond-the-Standard-Model effects of hadron-collider processes involving the interaction of two electroweak and two Higgs bosons, VVHH, with V being either a W or a Z boson.
We examine current experimental results by the CMS collaboration in the context of a dimension-8 extension of the Standard Model in an effective-field-theory formalism. We show that constraints from vector-boson-fusion Higgs-pair production on operators that modify the Standard Model VVHH interactions are already comparable with or more stringent than those quoted in the analysis of vector-boson-scattering final states. We study the modifications of such constraints when introducing unitarity bounds, and investigate the potential of new experimental final states, such as ZHH associated production. Finally, we show perspectives for the high-luminosity phase of the LHC.
}

\begin{document} 
\maketitle
\flushbottom

\section{Introduction}
\label{sec:intro}

Scattering processes involving the production of multiple electro-weak (EW) as well as Higgs (H) bosons are central to the entire physics programme of the Large Hadron Collider (LHC). Such reactions typically feature rich and complex signatures through which it is possible to probe several complementary properties of the underlying dynamics at the same time, allowing a detailed scrutiny of the gauge interactions they stem from. This can be exploited both for a precise extraction of fundamental Standard Model (SM) parameters, as well as to cast stringent bounds on potential beyond-the-Standard-Model (BSM) scenarios.

In the SM, the underlying $SU(2)_L\otimes U(1)_Y$ gauge symmetry fully dictates the pattern and relative intensity of the triple VVV$'$ and quartic VVV$'$V$'$ gauge interactions (with ${\text{V}^{(\prime)} = \text{W}^\pm, \text{Z}}$), as well as of the VVH and VVHH Higgs-gauge interactions. Experimental measurement of such trilinear and quartic pure-gauge and Higgs-gauge couplings, which we collectively denote with TCs and QCs respectively, is thus especially suited to gather insight on the nature of the EW symmetry-breaking mechanisms, and to unveil potential deviations from the SM couplings in the EW gauge sector \cite{Belanger:1992qh, Belanger:1992qi,Eboli:2003nq,Degrande:2012wf,Falkowski:2016cxu, Nordstrom:2018ceg, Arganda:2018ftn}.

BSM scenarios can be conveniently analysed in a model-independent manner by resorting to an effective-field-theory (EFT) framework, in which indirect New Physics effects are parametrised by including in the Lagrangian higher-dimension operators that modify the production pattern of SM particles. In the Standard Model Effective Field Theory (SMEFT) \cite{Buchmuller:1985jz,Grzadkowski:2010es}, operators affecting both TCs and QCs start at dimension 6 \cite{Buchmuller:1985jz,Hagiwara:1993ck,Gonzalez-Garcia:1999ije}, and their effects are commonly constrained at the LHC by measuring the production of two EW bosons \cite{Green:2016trm,CMS:2021icx,CMS:2021foa,CMS:2020mxy,CMS:2019ppl,ATLAS:2021jgw,ATLAS:2021ohb}, or the vector-boson-fusion production of EW bosons \cite{CMS:2019nep,CMS:2017dmo,ATLAS:2020nzk,ATLAS:2021ohb},
even if weaker constraints 
can also be obtained from more complex
processes such as vector-boson scattering (VBS)~\cite{Bellan:2021dcy}. In particular, in a dimension-6 linear EFT extension of the SM VVH and VVHH vertex factors are totally correlated. Modifications of QCs with no TC contamination are instead more subtle as, assuming EW symmetry is linearly realised, they only arise in operators of dimension 8 or higher, which are dubbed for this reason \emph{genuine} anomalous quartic operators~\cite{Eboli:2006wa,Almeida:2020ylr}. As a consequence, they mainly affect the production of three bosons or the VBS di-boson channels. Effects of Higgs compositeness, encoded at low energy in an extension of the SM like the Higgs EFT (HEFT)~\cite{Feruglio:1992wf,Grinstein:2007iv}, do also induce genuine QC variations, and their relation with dimension 8 or higher SMEFT operators has been studied in Refs.~\cite{Brivio:2013pma,Brivio:2016fzo}.

In this article we concentrate on the analysis of genuine SMEFT anomalous quartic operators, and set new constraints on their effects by means of a thorough investigation of the VVHH interaction\footnote{We neglect all contributions from SMEFT operators of dimension 6, in order to directly compare our results with the procedure adopted by CMS. A more refined analysis should consistently include constraints on dimension-6 operators, which is beyond the scope of our work.}. The production processes we consider are Higgs-pair production, both in vector-boson fusion mode (VBF-HH) and in association with a Z boson (ZHH), as well as the loop-induced production of a single Higgs in association with two Z bosons (\ggzzh).
We study these reactions by means of a simplified version of the analysis that the CMS experiment employs to constrain genuine QC operators in VBS. After validating this procedure against public CMS QC constraints ~\cite{CMS:2020gfh,CMS:2019qfk,CMS:2020ioi,CMS:2020ypo} in absence of unitarity bounds, we show that constraints from VBF-HH on genuine QC operators are comparable with or more stringent than those quoted in the analysis of the VBS signature. Similar investigations are performed on the ZHH and \ggzzh\ channels, both at current LHC luminosity as well as for the high-luminosity-LHC (HL-LHC) phase.
We do not consider in this study the q$\bar{\mathrm{q}}$-initiated ZZH process. At the LHC its cross section, known at NLO in the SM~\cite{Baglio:2015eon,Baglio:2016ofi}, is larger than that of the gluon-fusion process, however it is not within current data sensitivity and so no results from LHC collaborations exist. As our study constrains operators that specifically modify QCs without altering TCs, we only focus on processes exhibiting quartic vertices at the lowest order.

We choose the subset of possible dimension-8 operators which modify VVHH vertex strengths following Refs.~\cite{Eboli:2006wa,Almeida:2020ylr}. Among these operators, the ones containing no EW field strength tensor are categorised as \emph{scalar}, and denoted with a subscript `S', while \emph{mixed} operators (subscript `M') feature two covariant derivatives of the Higgs doublet and two field strengths. We do not consider \emph{transverse} (`T') operators in our analysis, as their effect on the VVHH vertex is found to be negligible, compatibly with the results of Ref.~\cite{Brass:2018hfw}.

By analysing data at a centre-of-mass energy of 13 TeV, the CMS collaboration has determined limits on the Wilson coefficients $f_X/\Lambda^4$ ($\Lambda$ being the New Physics scale) of the above operators from several VBS final states, which are identified by the presence of two vector bosons (V or $\gamma$) in the final state and two jets with a large invariant mass and rapidity difference. The ATLAS collaboration has performed several VBS measurements at 13 TeV as well, but it has not extracted limits on dimension-8 operators so far. A full review of VBS measurements can be found in Refs.~\cite{Rauch:2016pai,Covarelli:2021gyz} and a compilation of constraints on dimension-8 operators is available in~\cite{aQGCcomp}.
The most stringent limits on the coefficients of the dimension-8 operators \fszero, \fsone, \fmzero, \fmone, and \fmseven\ are derived from the simultaneous study of leptonic WZ and same-sign WW VBS~\cite{CMS:2020gfh}, and from semileptonic WV VBS searches, where the hadronic decays of W and Z are not resolved~\cite{CMS:2019qfk}.
The latter work also contains semileptonic ZV searches, which have weaker sensitivities to all operators. The most stringent limits quoted by the CMS collaboration on the remaining operators $f_{\mathrm{M}2-5}$ result from VBS W$\gamma$ and Z$\gamma$ analyses~\cite{CMS:2020ioi,CMS:2020ypo}. However,
we will show that such constraints can be improved by investigating these operators in the aforementioned VV analyses. The CMS collaboration does not quote any limit on the
coefficient \fstwo, while limits on $f_{\mathrm{M}6}$ do not add further information, as the corresponding operator was found to be redundant~\cite{Perez:2018kav}.
Concerning multi-Higgs-boson final states, both the ATLAS and CMS collaborations have searched for Higgs-pair production in the VBF-HH mode, in the $2\mathrm{b2\overline{b}}$2j, $\mathrm{b\overline{b}}\tau^+\tau^-$2j, and $\mathrm{b\overline{b}}$2$\gamma$2j channels~\cite{CMS:2021ssj,CMS:2020tkr,CMS:2022pga,ATLAS:2020jgy}.
In these studies, BSM effects are described in a different parameterisation which is equivalent to inserting an effective VVHH vertex modifier \kappav\, where the SM corresponds to $\kappav = 1$~\cite{Bishara:2016kjn}. 
For this study, we consider the constraint $-0.1 < \kappav <$ 2.2 at the 95\% confidence level (@95\% CL) set by the CMS $2\mathrm{b2\overline{b}}$2j analysis~\cite{CMS:2021ssj}.\footnote{During the review process of this article, a new set of results stemming from a combination
of different final states has become available~\cite{CMS:2022dwd} that supersedes this limit; we point out that this does not undermine the validity of the methodologies and results shown in this paper.}
For other tri-boson final states that we will consider in this work no experimental studies exist. In this work, approximate estimates of the sensitivities are performed for the ZZH and ZHH final states only, since tri-boson processes containing simultaneously W$^\pm$ bosons and ${\text{H}\rightarrow \mathrm{b\overline{b}}}$ decays are contaminated by large-rate $\mathrm{t\overline{t}}$(+X) backgrounds and cannot therefore be estimated correctly in a simplified analysis.

The typical effect of QC modifications is to distort differential spectra, primarily the invariant mass of the produced multi-boson system (e.g. $\mvv$ for di-bosons), with respect to the SM baseline, leading to enhanced rates in the high-energy tails. By increasing QCs, scattering amplitudes grow up to the point in which they violate unitarity bounds, signalling the breakdown of a truncated EFT approach, and the necessity of including explicit New Physics degrees of freedom in the Lagrangian. Unitarity constraints based on a partial-wave decomposition of matrix elements \cite{Almeida:2020ylr} are then important to guarantee the reliability of the exclusion bounds cast by the experimental searches, although CMS EFT-based analyses do not follow a uniform prescription to enforce them. In particular in~\cite{CMS:2019qfk} unitarity effects are neglected, and the limits are set using events with arbitrarily large values of $\mvv$. In~\cite{CMS:2020ioi,CMS:2020ypo} no upper $\mvv$ bound or signal hypotheses are applied on the data, but the operator-dependent unitarity limits of \cite{Almeida:2020ylr} are computed \emph{a posteriori} based on the measured limits on Wilson coefficients: since the signal model extends however beyond this limit, this may lead to a circular argument. Finally, in~\cite{CMS:2020gfh} physical limits are obtained by cutting the EFT contribution off at the unitarity bound, and removing all but the expected SM contribution for higher values of $\mvv$ (`clipping' method~\cite{Contino:2016jqw}). This approach is also questionable, as the signal model features an unphysical sharp drop at the unitarity bound, which depends on the value of the Wilson coefficient under examination. Also, more importantly, experimental data beyond the bound are still used to set the limit, stretching the EFT approach outside its validity domain. Studies on the uncertainty related to the unitarisation methods are also available~\cite{Garcia-Garcia:2019oig}.

In this article we adopt a dedicated technique to incorporate unitarity effects on EFT operators in a physically consistent manner, similarly to what done in Refs.~\cite{Kalinowski:2018oxd,Kozow:2019txg,Chaudhary:2019aim}. First, limits on a given EFT operator are set as functions of $\mvv$ neglecting unitarity, by using the simulated cumulative $\mvv$ distributions (as opposed to total cross sections within fiducial cuts) for the analysed channels. The obtained $\mvv$-dependent exclusion is then compared to the unitarity constraints for that operator, which are naturally in the form $|f_X| < f_{\max,X}(\mvv)$, and the best attainable
limits are defined by considering the experimental data in the maximum $\mvv$ range in which the effects of the operator do not violate unitarity.
\\

The structure of the paper is as follows: in section \ref{sec:simu} we introduce the employed inputs from experimental analyses and the setup of our simulations, with particular reference to the inclusion of unitarity bounds; sections \ref{sec:vali} and \ref{sec:unit} show the validation against the CMS analysis of VBS; in sections \ref{sec:vbfhh} and \ref{sec:zzhzhh} results are presented relevant for the VBF-HH, ZHH, and \ggzzh\ channels at the LHC, with projections for HL-LHC detailed in section \ref{sec:hllhc}; we finally conclude in section \ref{sec:summ}.

\section{Simulation setup and observables}
\label{sec:simu}

The simulation of the final states under examination is performed
using version 2.7.3 of the \mgamc\ event generator~\cite{Alwall:2014hca, Frederix:2018nkq} at leading order (LO) in QCD\footnote{\mgamc\ allows to simulate the relevant SM background processes at NLO accuracy, as documented in Refs.~\cite{Frederix:2011qg,Frederix:2014hta,Bagnaschi:2018dnh,Ballestrero:2018anz}. However, we consider LO simulations for compatibility with the available BSM models, and in keeping with the CMS Monte Carlo generation setup.} at the parton level.
LHC conditions are reproduced using symmetric 
proton-proton collisions at $\sqrt{s}= 13$ TeV.
We consider the following processes which are sensitive to dimension-8 EFT effects:
\begin{enumerate}
\setlength\itemsep{0em}
    \item EW production of two same-sign W bosons and two quark jets (W$^\pm$W$^\pm$ VBS);
    \item EW production of a W and a Z boson, associated with two quark jets (W$^\pm$Z VBS);
    \item EW production of two opposite-sign W bosons and two quark jets (W$^+$W$^-$ VBS);
    \item EW production of two Higgs bosons and two quark jets (VBF-HH);
    \item EW production of two Higgs and a Z boson (ZHH);
    \item Loop-induced (LI) production of two Z and a Higgs boson from a gluon-gluon initial state (\ggzzh).
\end{enumerate}

\begin{figure}[htbp]
\centering
\includegraphics[width=.9\textwidth]{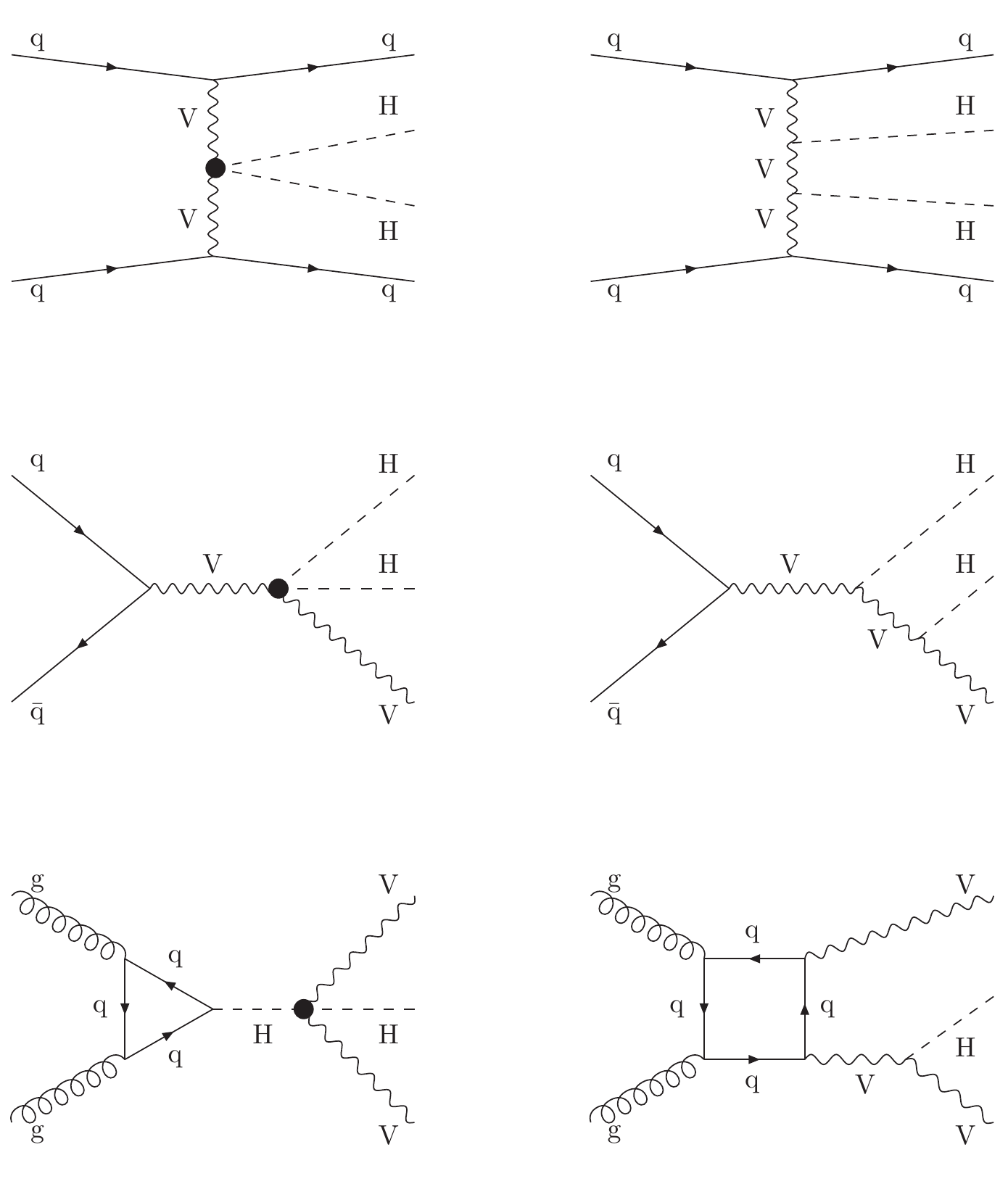} \caption{Sample diagrams relevant to the VBS channel (top row), the VHH channel, (middle row), and the loop-induced ggVVH channel (bottom row). Contributions from higher dimension operators are collected on the left, while SM interactions are on the right.}
\label{fig:diagrams}  
\end{figure}

Figure~\ref{fig:diagrams} shows typical
lowest-order diagrams for the processes
outlined above, with (left) and without
(right) EFT insertions. Since signal process 5 
has not been searched for at the 
LHC so far, we also simulate the main expected background for this process when the dominant H $\rightarrow \mathrm{b\overline{b}}$
decay channel is chosen to reconstruct Higgs bosons:
\begin{enumerate}
\setlength\itemsep{0em}   
\setcounter{enumi}{6}
\item QCD-mediated production of a Z boson with two b and two anti-b quarks.
\end{enumerate}

Table~\ref{tab:processes} lists the specific parameters used to generate the aforementioned processes. The number of active flavours in protons and jets (4 or 5) is chosen according to whether the related experimental analysis does not or does apply a b-jet veto on tagging jets. All simulations share the same choice
for SM EW parameters:
\begin{eqnarray}
1/\alpha_{\mathrm{EW}} &=& 127.9, ~~~~~~~~ G_F = 1.166377 \cdot 10^{-5}~\mathrm{GeV}^{-2}, \nonumber \\
m_{\mathrm{H}} &=& 125~\mathrm{GeV}, ~~~~  m_{\mathrm{Z}} = 91.1876~\mathrm{GeV},   \nonumber \\
m_{\mathrm{t}} &=& 172~\mathrm{GeV},  ~~~~  V_{\mathrm{CKM}} = {\mathds 1}_3.
\end{eqnarray}
The parton distribution functions used in the simulations are taken from the NNPDF2.3 LO set~\cite{Ball:2013hta} with $\alpha_s(m_{\rm Z}) = 0.130$.
\begin{table}[t] {
  \centering
  \begin{tabular}{|c|c|c|c|c|c|}
\hline
Process & \mgamc & QCD & Max. & CMS & \mysigma[1.1, 13 TeV] \\
number & syntax & order & jet flav. & result & SM (fb) \\
\hline
\multicolumn{6}{|c|}{Signal (including EFT effects)} \\
\hline
1 & \texttt{p p > w+ w+ j j QCD=0} & LO & 4 & \cite{CMS:2020gfh,CMS:2019qfk} & 4.514(9) \\
& \texttt{p p > w- w- j j QCD=0} & & & & \\
2 & \texttt{p p > w+ z j j QCD=0} & LO & 4 & \cite{CMS:2020gfh,CMS:2019qfk} & 8.55(2) \\
& \texttt{p p > w- z j j QCD=0} & & & & \\
3 & \texttt{p p > w+ w- j j QCD=0} & LO & 4 & \cite{CMS:2019qfk} & 9.97(2) \\
4 & \texttt{p p > h h j j QCD=0} & LO & 5 & \cite{CMS:2021ssj} & 0.0329(7) \\
5 & \texttt{p p > z h h QED=3} & LO & 5 & - & $0.01295(5)$ \\
6 & \texttt{g g > z z h [noborn=QCD]} & LI (LO) & 5 & - & $3.493(7)\times10^{-3}$ \\
\hline
\multicolumn{6}{|c|}{Background (SM only)} \\
\hline
7 & \texttt{p p > z b b\~ ~b b\~} & LO & 4 & - & $0.729(3)$ \\
\hline
  \end{tabular}
  \caption{LHC processes of interest studied in the present work, numbered according to the main text. The number of active flavours considered in protons and jets is also listed. $\mysigma[m_{\min},m_{\max}]$ is the cross section in the interval $m_{\min}\le m\le m_{\max}$, with $m$ being the invariant mass of the produced di- or tri-boson final state. Each cross-section value is evaluated
  after cuts on final-state jets described in the text and reports in brackets the integration uncertainty on the last digit. 
  }
  \label{tab:processes} 
  }
\end{table}

EFT effects are simulated by loading in \mgamc\ suitable UFO models~\cite{Degrande:2011ua}, containing the relevant modifications to SM vertex structures. For all signal LO simulations, the latest
models available at~\cite{twikiEboli} were used. It must be noticed
that these differ in several aspects from those used in the 
CMS analyses~\cite{CMS:2020gfh,CMS:2019qfk}, which date back to  2013. Taking into consideration observations in 
recent papers such as~\cite{Perez:2018kav}, the list and the content of EFT operators were revised by removing redundant operators (e.g.~M6), redefining others (e.g.~M5) and adding new ones (e.g.~S2), in order to obtain a complete basis.

For the loop-induced \ggzzh\ simulation, a dedicated UFO model has been constructed, 
since the original one did not support loop diagrams. 
Starting from a NLO QCD description of the SM, new Lorentz
structures corresponding to the additional operators were
inserted. This was possible since none of these operators involves coloured particles. While
in this paper such a model has been used only for the loop-induced process \ggzzh, it could be exploited
to refine our analysis including NLO QCD corrections for all processes, which is left for future work.
In the VBF-HH samples, modifications in terms of \kappav\ were instead performed by
changing the WWHH and ZZHH couplings by the same scaling
factor.

Our simulations include:
\begin{itemize}
    \item for processes 1-3 (VBS) and 5-6 (ZHH, \ggzzh), sets
    of samples with one Wilson coefficient $f_X$ varied as $f_X/\Lambda^4 = \{0, \pm 2, \pm 5, \pm 10, \pm 20\}$ TeV$^{-4}$ ($X = $ S{0-2}, M{0-7});
    \item for process 4 (VBF-HH), sets
    of samples with coefficient values as above, and additionally samples with 
    an effective vertex factor $\kappav = 0$, $1$, 
    $\pm 2$, $\pm 5$, $\pm 10$;
    \item for process 7 (background), just the 
    SM generation.
\end{itemize}

Since the EFT-sensitive region is located at 
high scattering energies, the decays of 
vector and Higgs bosons are expected to produce secondary
particles with large transverse momentum ($\pt \gg m_{\mathrm{H}}/2, m_{\mathrm{V}}/2$). For this reason we
do not apply parton showering to parton-level events, assuming
modifications to the transverse momentum \pt\ to be moderate.
Following the same reasoning, no selections are applied to 
the decay products of the gauge and Higgs bosons, since we assume that at large \pt\ the detector acceptance be constant over the considered phase space (and similar for signals and
physical backgrounds): since the experimental inputs already have acceptance effects folded in, these factorise in our approach. The only exception is for processes 5 and 7, where no experimental analysis exists and the approach followed in this case is discussed in Section \ref{sec:zzhzhh}. 

In the VBS and VBF processes (1-4), on the other hand,
typical experimental selections on the additional jets
are applied:
\begin{eqnarray}
\pt(\mathrm{j}) &>& 40~\mathrm{GeV}, ~~~~~~~~  |\eta(\mathrm{j})| < 4.7,\nonumber \\
m_{\mathrm{jj}} &>& 500~\mathrm{GeV}, ~~~~~~~ |\Delta\eta_{\mathrm{jj}}| > 2.5.
\end{eqnarray}
In the background process (7)
a cut on pairs of b-jets is
applied to emulate a possible
experimental selection. It is required that the invariant mass of at least two of the possible $\text{b}\bar{\text{b}}$ pairs, which can be obtained from combining the four b-jets, satisfies the requirement 115 GeV $< m_{\text{b}\bar{\text{b}}} <$ 135 GeV.

\begin{figure}[htbp]
  \centering 
  \includegraphics[width=.46\textwidth]{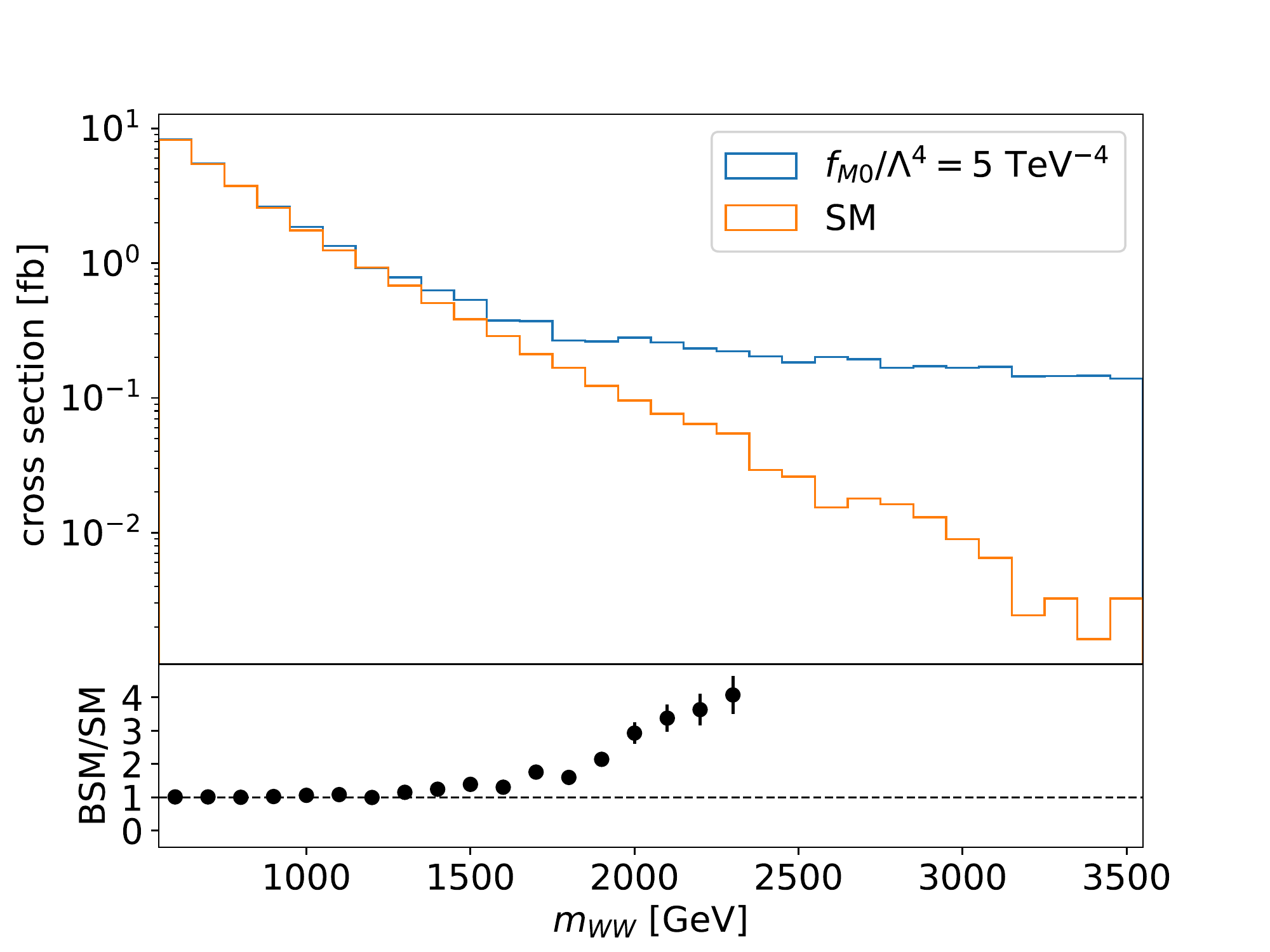} 
  \includegraphics[width=.46\textwidth]{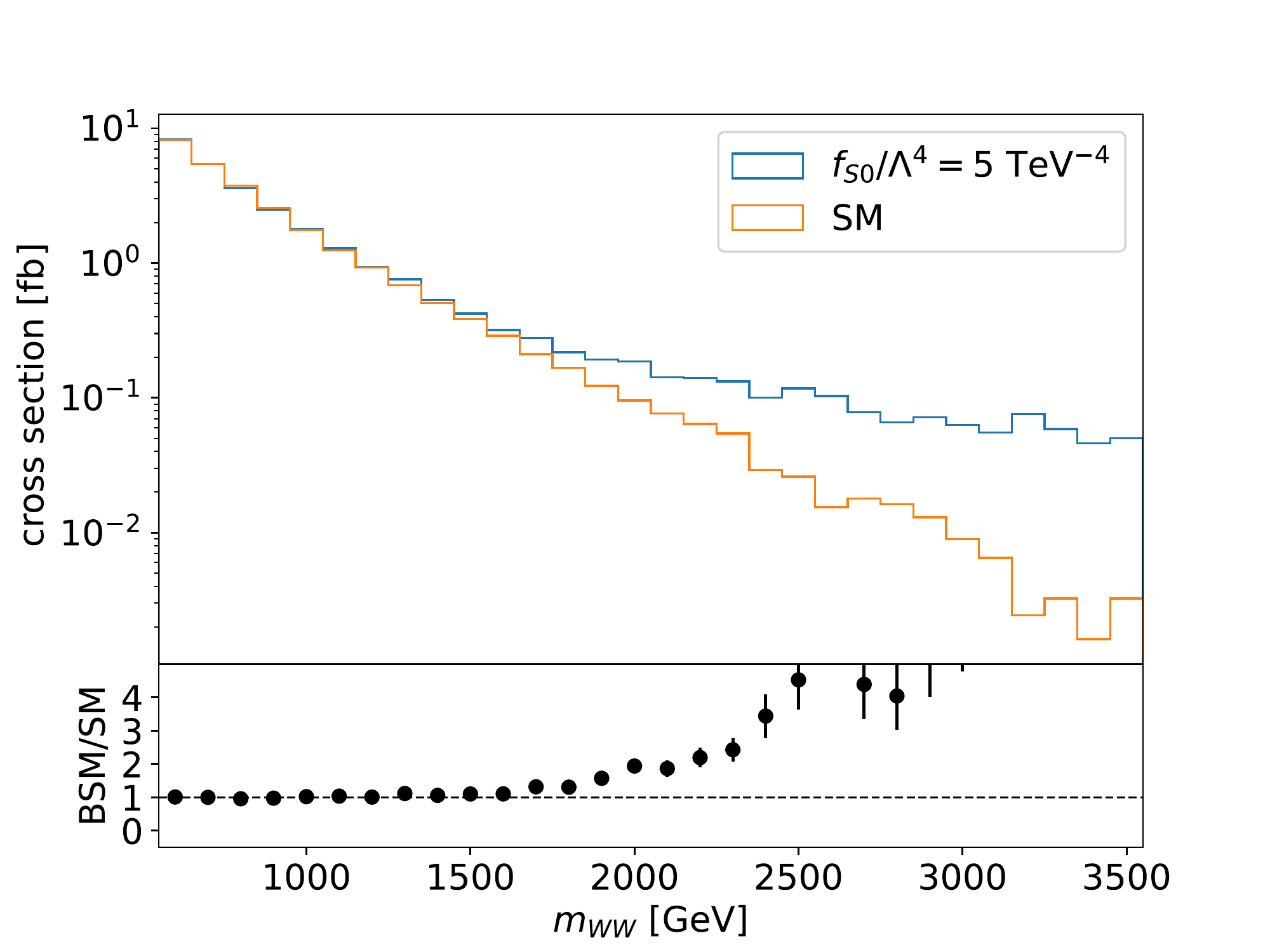} \\
  \includegraphics[width=.46\textwidth]{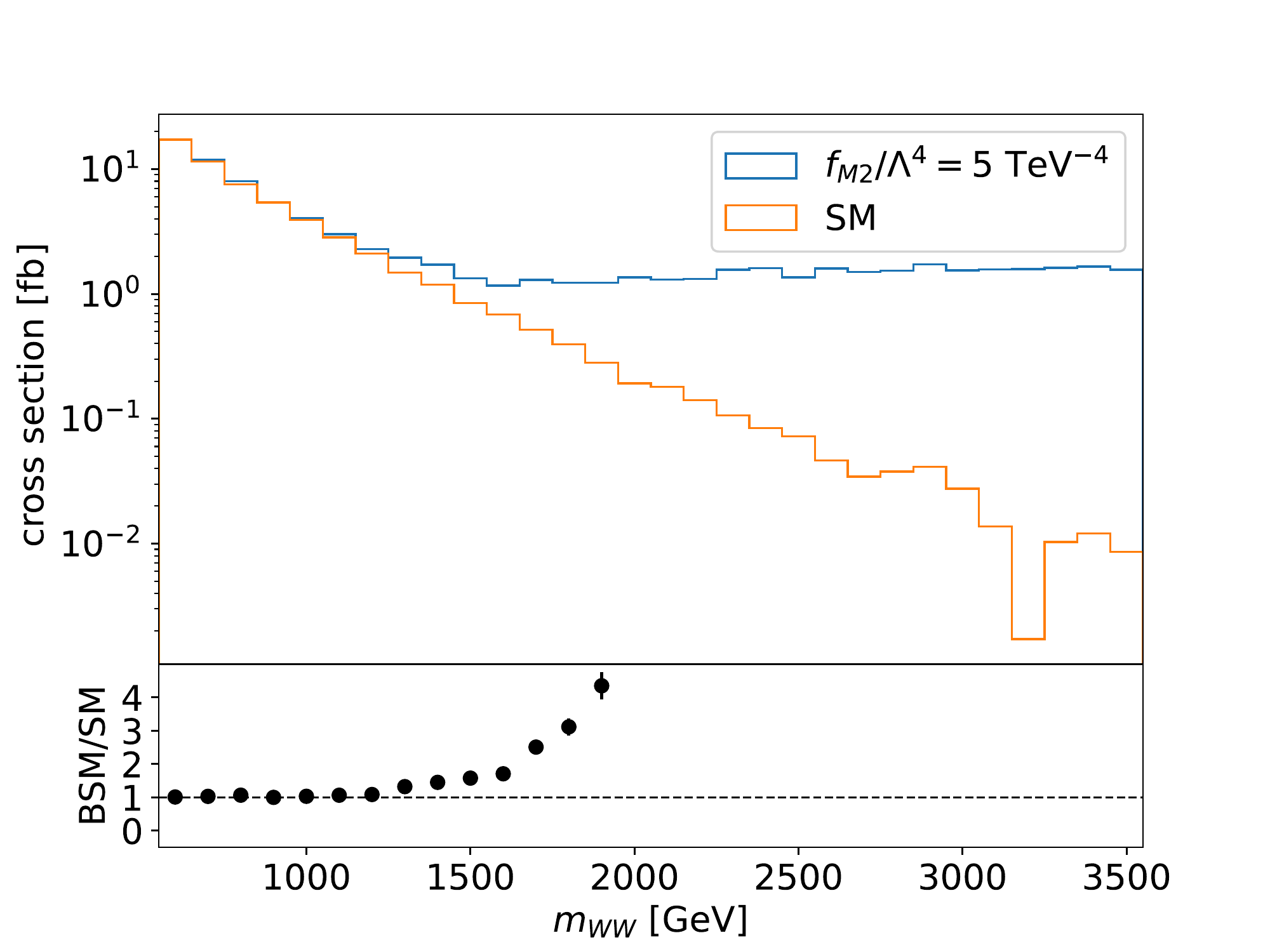} 
   \includegraphics[width=.46\textwidth]{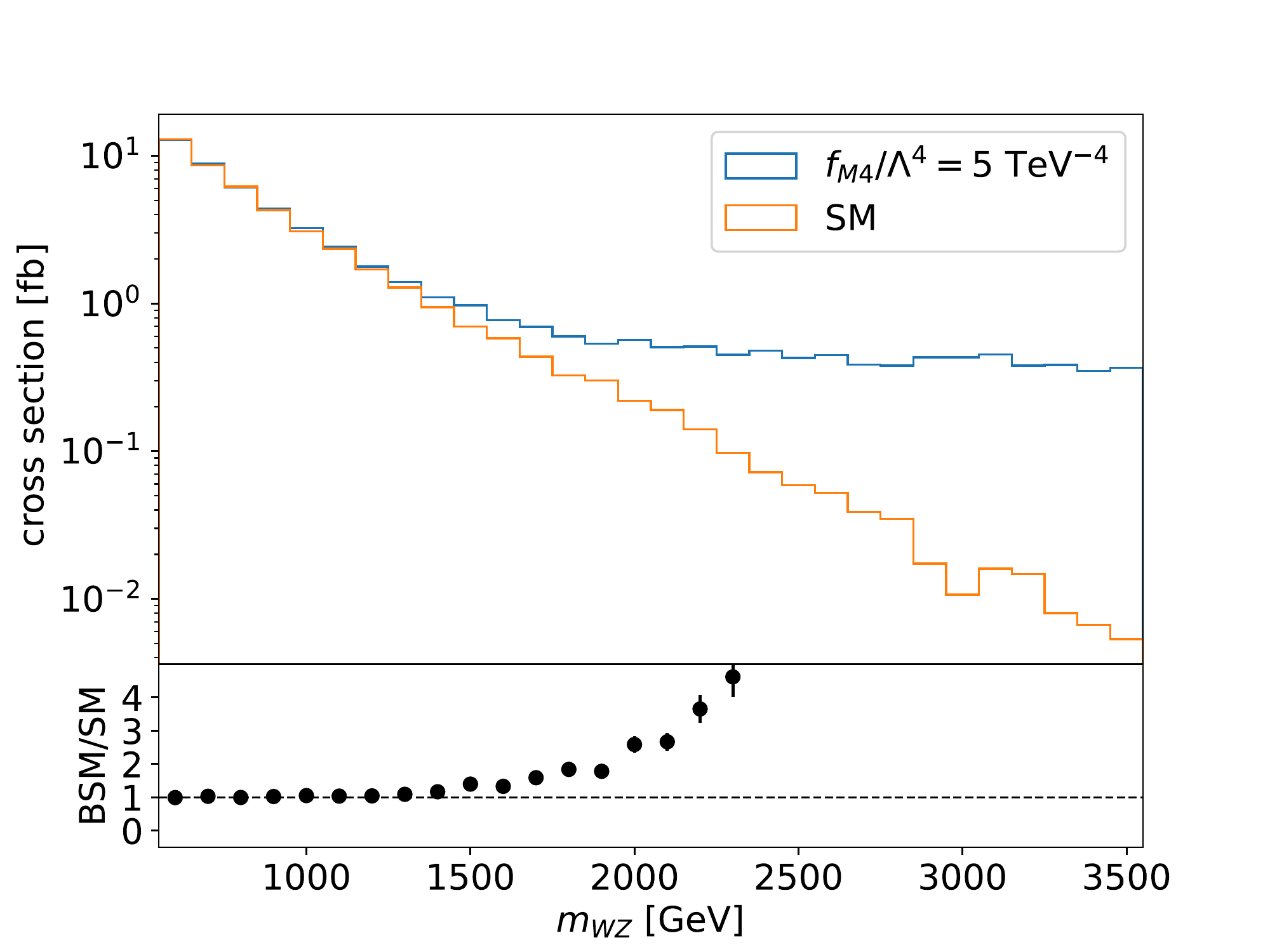}
   \caption{Examples of differential cross sections highlighting the high invariant-mass region for the processes VBS W$^\pm$W$^\pm$ (top left/top right), VBS W$^+$W$^-$ (bottom left) and VBS W$^\pm$Z (bottom right). The SM hypothesis is shown, as well as that of a single
   EFT operator (listed in the legend) with a nonzero
   coefficient $f_X/\Lambda^4$ set to 5 TeV$^{-4}$.
   EFT-over-SM ratio plots are presented in the lower panels. }\label{fig:vbs_mvv}  
\end{figure}

\begin{figure}[htbp]
  \centering 
  \includegraphics[width=.46\textwidth]{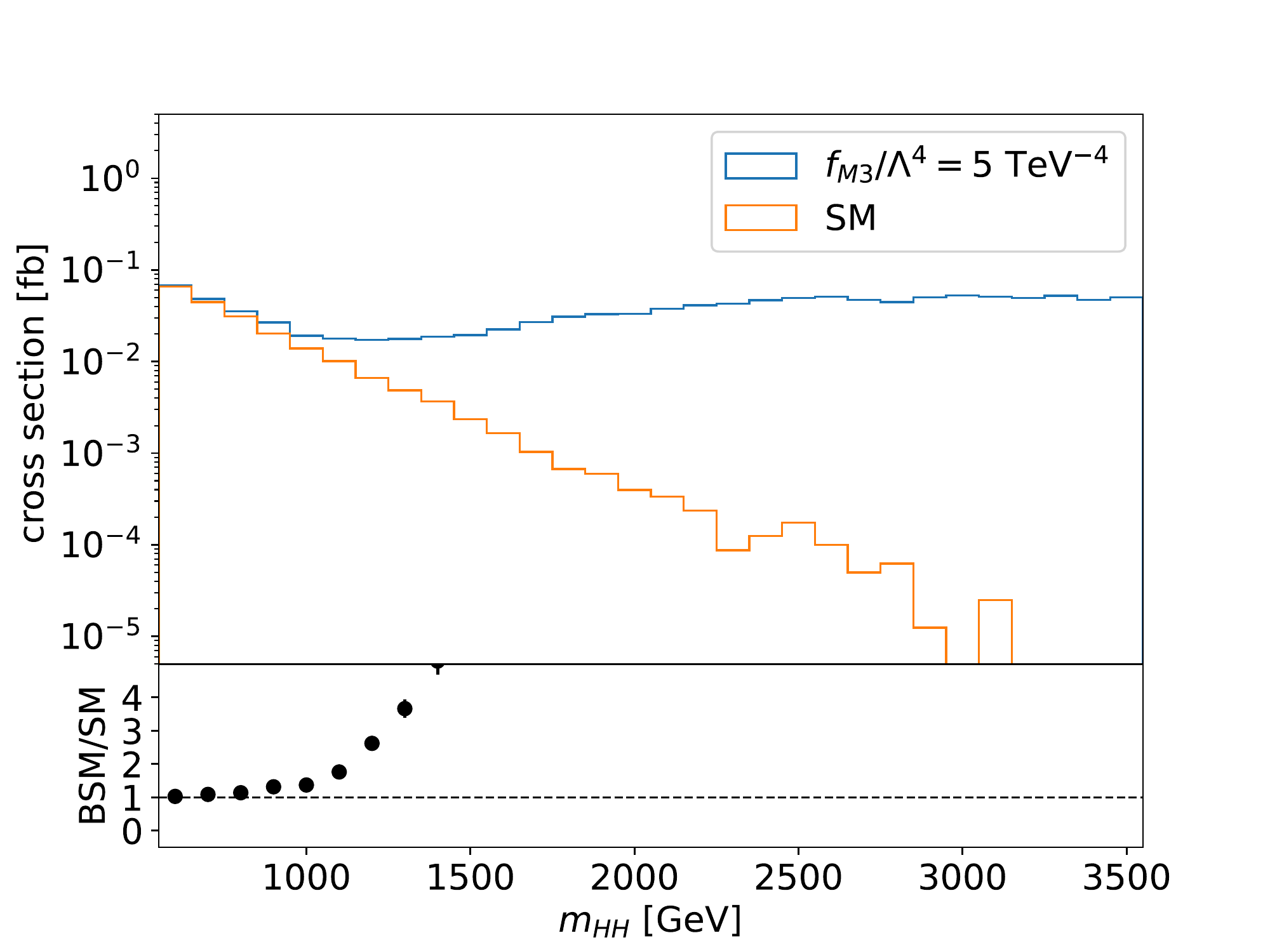}
  \includegraphics[width=.46\textwidth]{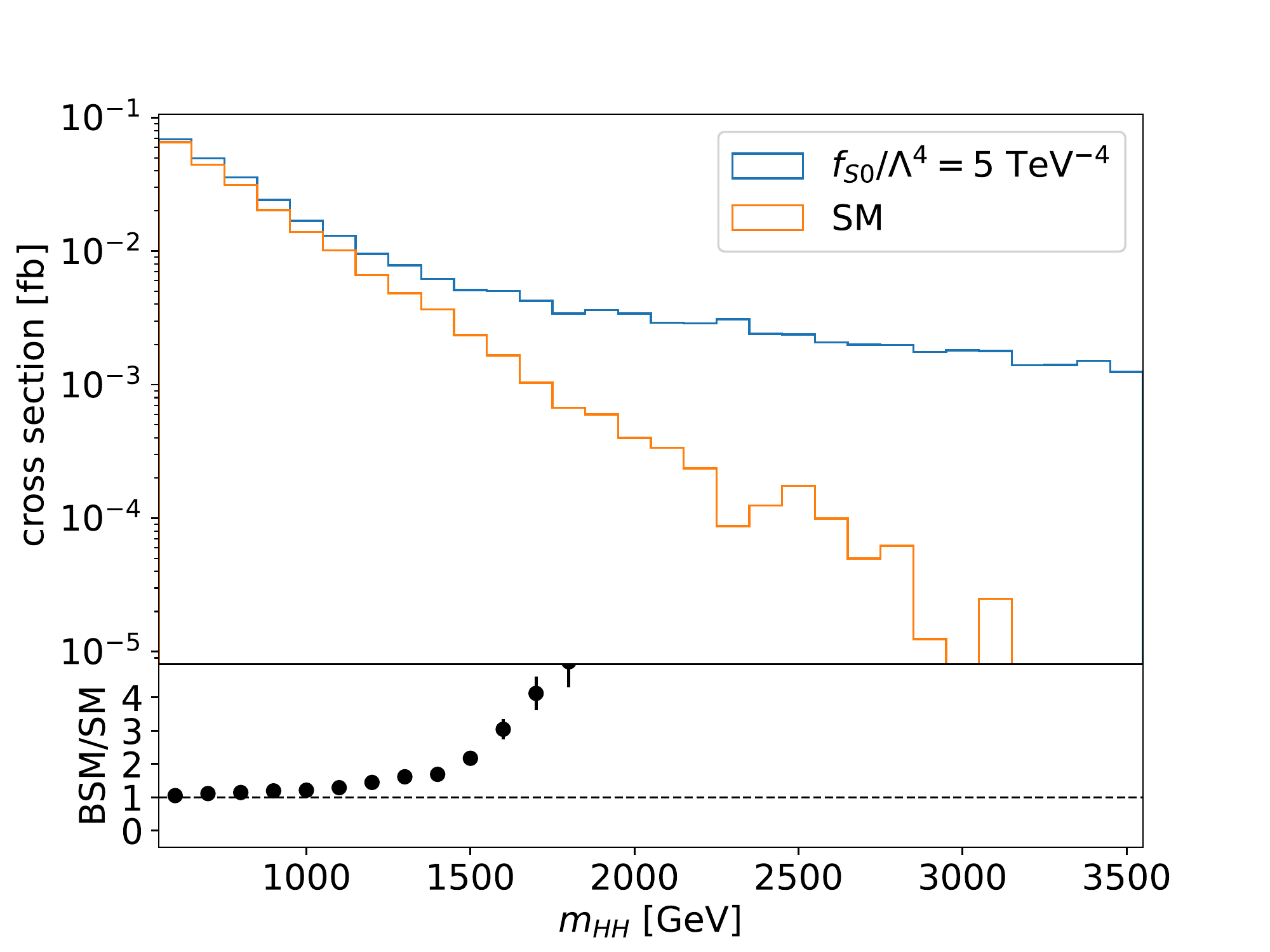} \\
  \includegraphics[width=.46\textwidth]{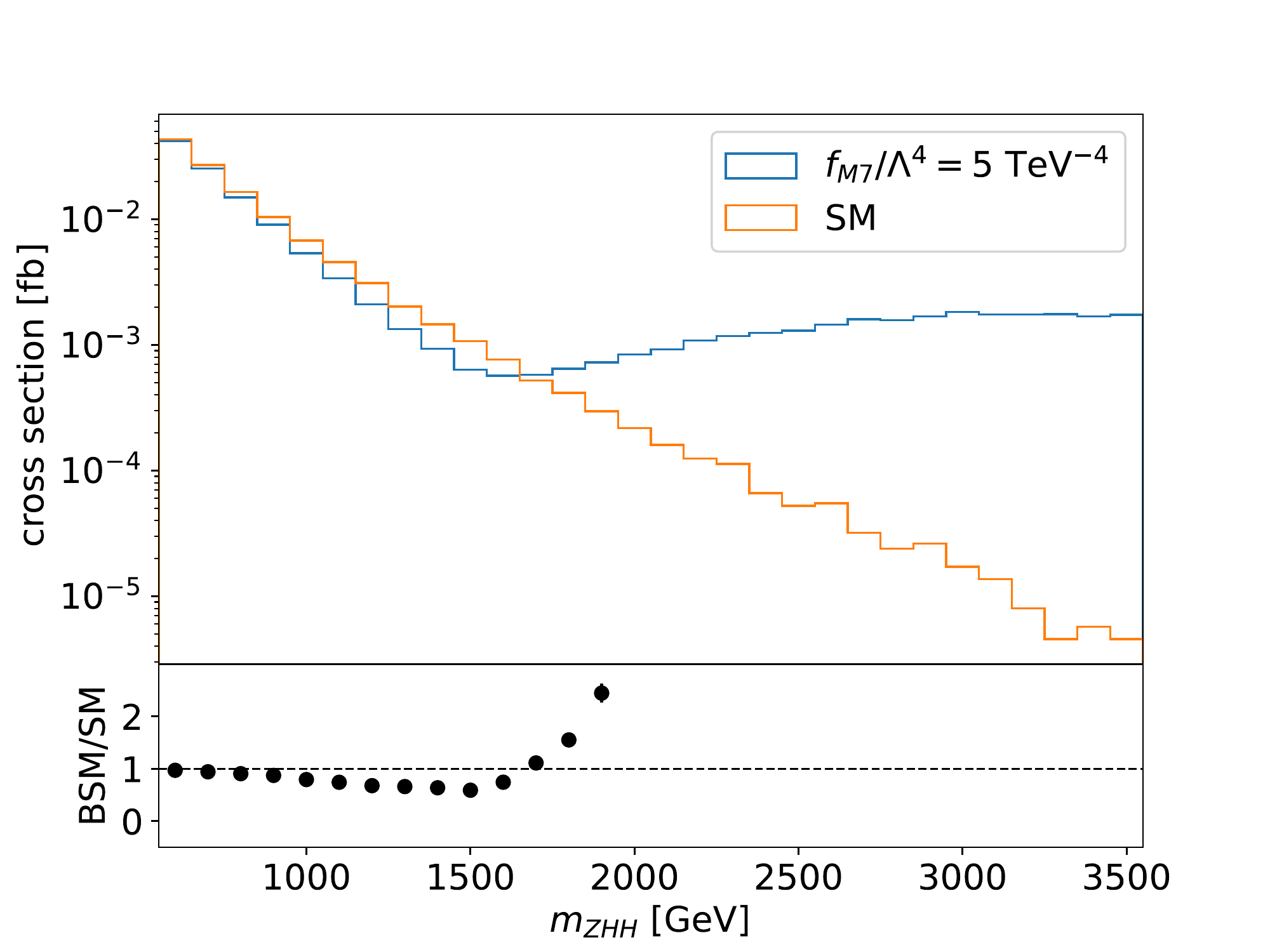}
  \includegraphics[width=.46\textwidth]{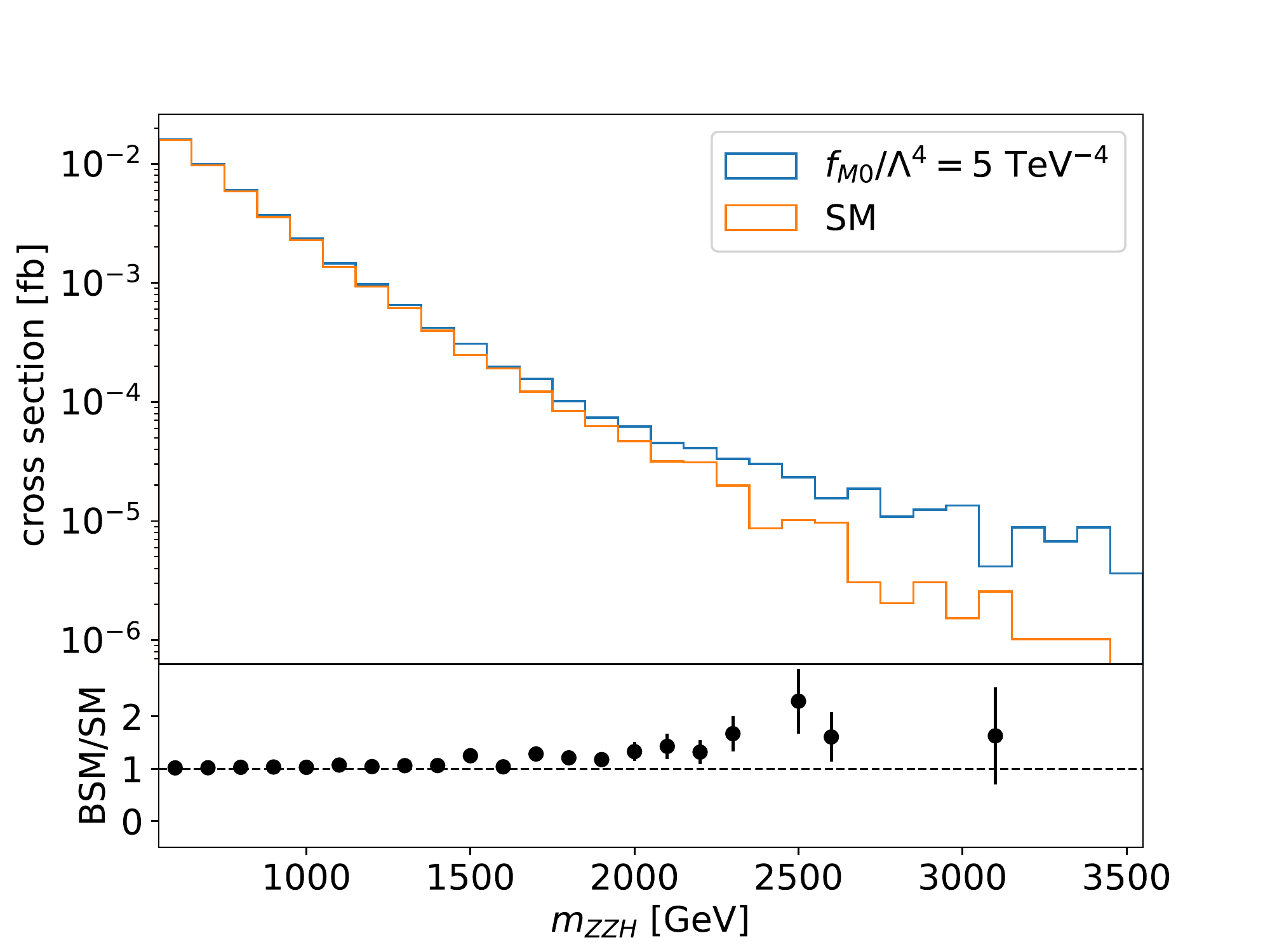}
   \caption{Examples of differential cross sections highlighting the high invariant-mass region for the processes VBF-HH (top left/top right), ZHH (bottom left) and \ggzzh\ (bottom right). The SM hypothesis is shown, as well as that of a single
   EFT operator (listed in the legend) with a nonzero
   coefficient $f_X/\Lambda^4$ set to 5 TeV$^{-4}$.
   EFT-over-SM ratio plots are presented in the lower panels.}\label{fig:hh_mhh}  
\end{figure}

Examples of simulated di-boson or tri-boson invariant-mass
spectra are shown in Figs.~\ref{fig:vbs_mvv} and \ref{fig:hh_mhh}.
We notice that, while final states with Higgs
bosons have SM cross sections which are a few
orders of magnitude smaller with respect to VBS, EFT
effects which affect tree-level amplitudes induce very large
enhancements at high invariant mass, especially in the case of mixed M-type operators. In all cases the ratios BSM/SM in the lower panels of Figs.~\ref{fig:vbs_mvv} and \ref{fig:hh_mhh} tend to unity at low invariant mass: this is an indirect confirmation that the modifications induced by dimension-8 operators on SM structures, like EW boson propagators, are numerically negligible for the purpose of our analysis.

CMS uses Monte Carlo templates to
perform binned data analyses and extract limits on the Wilson coefficients of dimension-8
operators. Typical data distributions used with this aim
are the di-boson invariant mass, or
the transverse mass if the final state contains one or
more undetected neutrinos whose transverse momenta are inferred
from the event imbalance in the transverse plane. This 
procedure requires a full description of the
SM backgrounds as well as a precise simulation of detector
effects. In order to obtain a simplified estimate of the
relative EFT sensitivities for different final states, we
use as a physical observable the cross section (or equivalently
the expected number of observed events at given integrated luminosity) in the interval $m_{\min} \le m \le m_{\max}$, with $m$ being the invariant mass of the produced di- or tri-boson states in the various channels; we denote such a cross section as $\mysigma[m_{\min}, m_{\max}]$, or simply as $\mysigma$.

Unitarity bounds for all operators are computed as
functions of $f_X/\Lambda^4$ using Ref.~\cite{Almeida:2020ylr}, resulting in more
stringent bounds than those provided by the VBFNLO
program~\cite{Arnold:2008rz}, which only considers $s$-wave scattering.
In all analyses presented in the following, $m_{\min}$ is fixed to a default value of 1.1
TeV. The choice is a compromise between defining an
EFT-enriched region where possible SM backgrounds
are negligible, and having a $m_{\min}$ value
below the unitarity bound in the
whole range of Wilson coefficients considered in the
analysis. The values of \mysigma\ using
$m_{\min} = 1.1$ TeV and $m_{\max} = \sqrt{s}$ (namely with no upper bound) are shown in the
rightmost column of Table~\ref{tab:processes}.
Repeating the analysis with $m_{\mathrm{min}}$ 
shifted by $\pm 100$ GeV
modifies the outcome of the validation
performed in Section~\ref{sec:vali} by less than 10\%,
for all operators.

\section{Validation on VBS without unitarity constraints}
\label{sec:vali}

As a first step we show that, although experimental constraints result from template analyses of the observable final states, constraints purely based on $\mysigma[m_{\min}, m_{\max}]$ are a reliable proxy of the actual EFT sensitivity.
Since most CMS results do not use unitarity regularisation,
in this validation phase we consider $m_{\max} = \sqrt{s}$,
i.e.~no upper bound on the system invariant mass.

The \mysigma\ values are computed for each process and simulated sample. The corresponding cross sections for a given 
experimental final state are obtained by
multiplying \mysigma\ by the branching
fractions ($BF$) of the gauge and Higgs bosons involved. For instance, the cross section for the $2\mathrm{b2\overline{b}}$2j final state considered in~\cite{CMS:2021ssj} 
is obtained as $\mysigma($VBF-HH$) \times BF^2($H$ \rightarrow \mathrm{b \overline{b}})$. For semileptonic WV VBS~\cite{CMS:2019qfk}, the $BF$-weighted cross sections
of VBS W$^+$W$^-$, VBS W$^\pm$Z and VBS W$^\pm$W$^\pm$ 
are summed, as they all contribute to the observed final state. For the \ggzzh\ and ZHH processes, the
H$ \rightarrow \mathrm{b \overline{b}}$ mode is chosen, while
only Z decays to electrons and muons
are considered, since hadronic decays
are more challenging experimentally
and beyond the scope of this analysis.

\begin{figure}[t]
  \centering 
  \includegraphics[width=.49\textwidth]{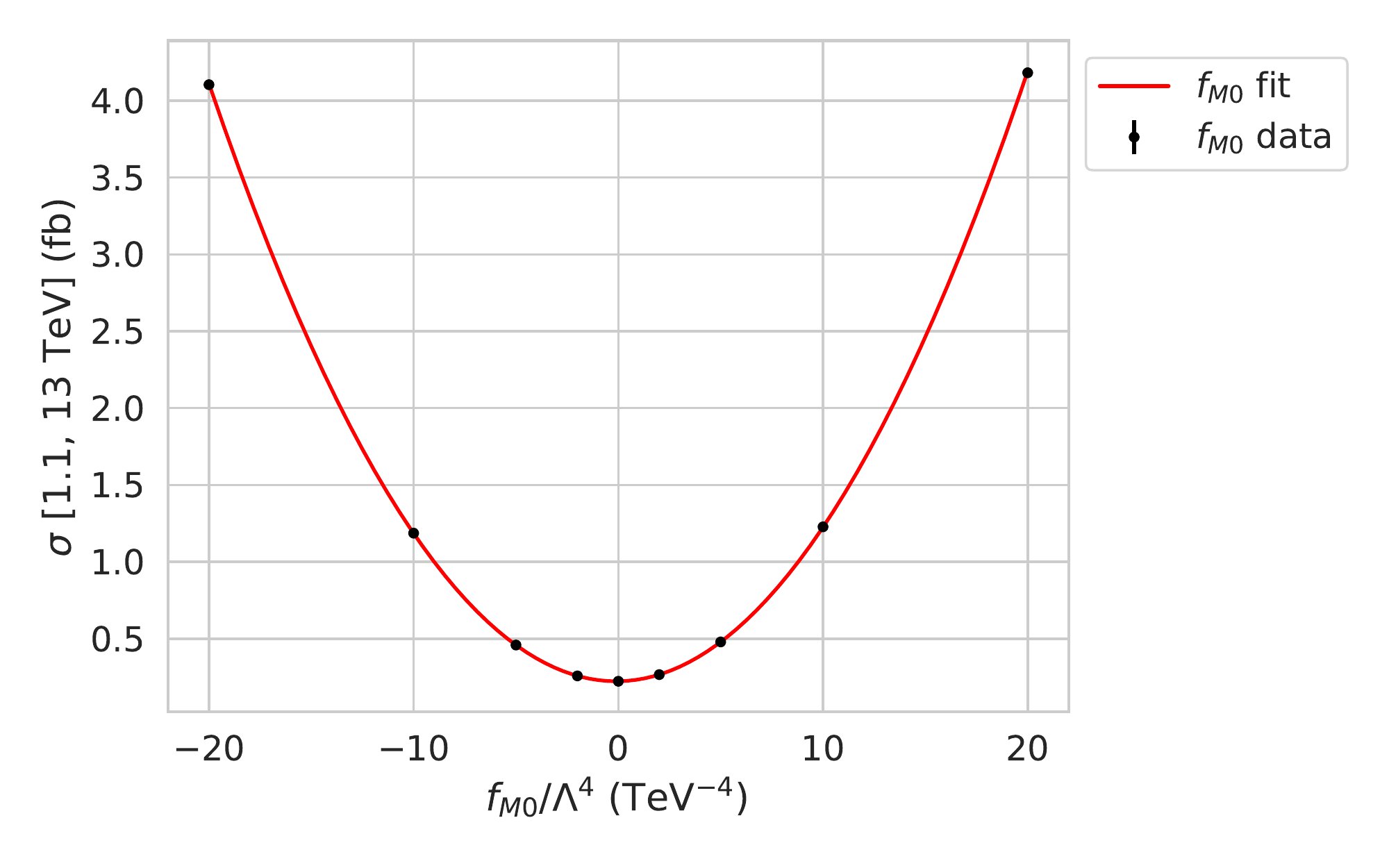}
  \includegraphics[width=.49\textwidth]{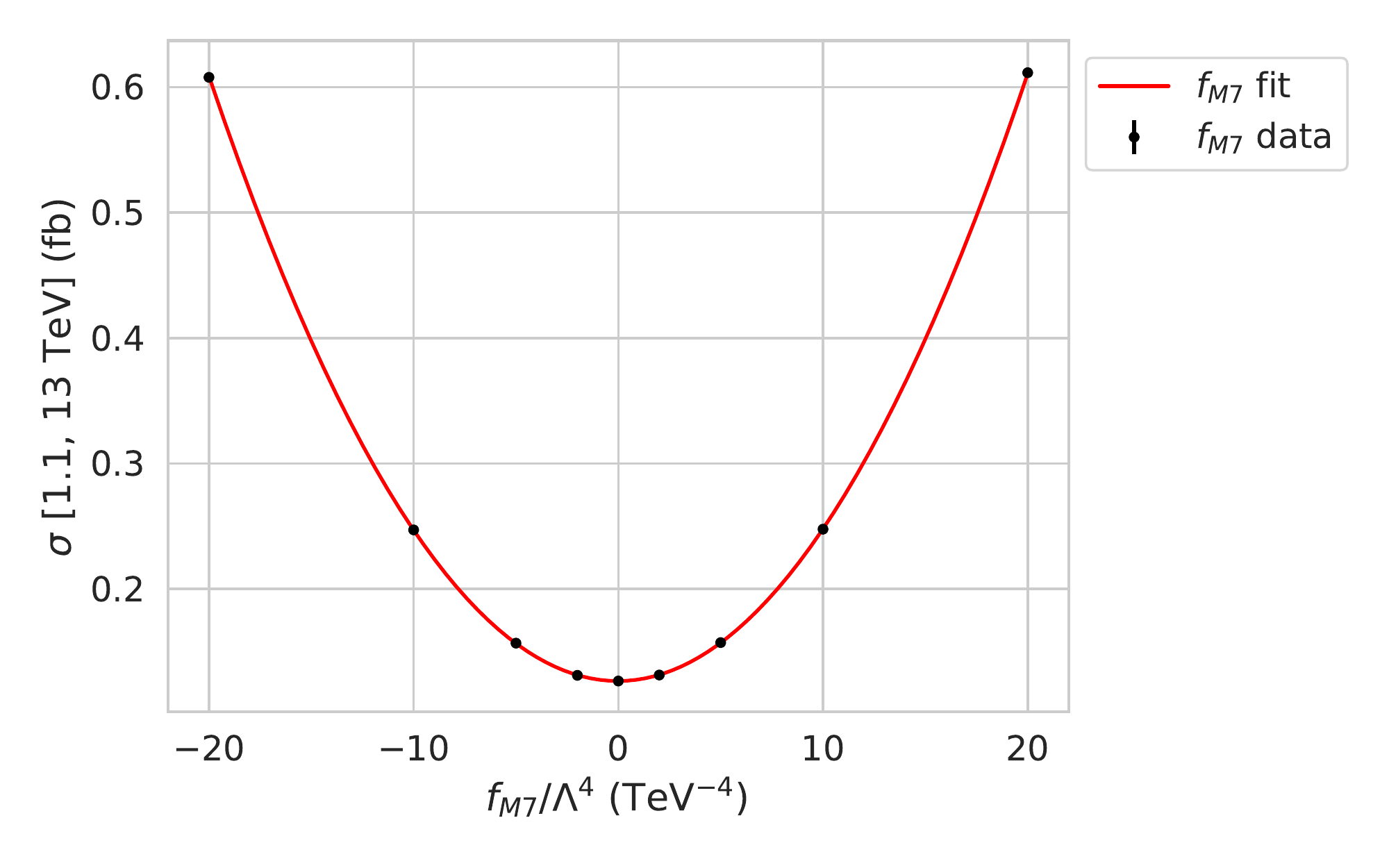} 
   \caption{Examples of quadratic fits of \mysigma[1.1 TeV, 13 TeV] varying the coefficient of a single nonzero dimension-8 operator. Left: \fmzero\ fit for the VBS W$^\pm$W$^\pm \rightarrow 2 \ell 2 \nu$ process. Right: \fmseven\ fit for the VBS W$^\pm$Z$ \rightarrow 3 \ell \nu$ process.} \label{fig:fits}  
\end{figure}

Cross sections for each channel are computed as functions of $f_X/\Lambda^4$, and quadratic fits are performed on the obtained results. Examples of such parabolic fits are shown in Fig.~\ref{fig:fits} for two of the processes under consideration. The validation proceeds through the steps below. 
\begin{enumerate}
\setlength\itemsep{0em} 
    \item A CMS experimental analysis is selected, and the published 95\% CL exclusion limit on a randomly-chosen operator coefficient is considered. 
    \item The exclusion limit is imposed on the parabola corresponding to the chosen operator, yielding a
    95\% CL exclusion limit on \mysigma. 
    \item Upon applying such a \mysigma-based exclusion on the parabolae corresponding to all other operators, exclusion limits on the corresponding coefficients are in turn 
    determined.
    \item As a closure test, the limits obtained with this procedure are systematically compared with the CMS published ones.
    \item Steps 1-4 are then repeated for different choices of the initial input operator, in order to check that similar bounds are obtained.
\end{enumerate}

\begin{table}[t] {
  \centering
  \begin{tabular}{|c|c|c|c|c|c|c|}
\hline
& \multicolumn{2}{|c|}{VBS W$^\pm$W$^\pm \rightarrow 2\ell 2\nu$} & \multicolumn{2}{|c|}{VBS W$^\pm$Z $\rightarrow 3\ell\nu$} &  \multicolumn{2}{|c|}{VBS W$^\pm$V semileptonic} \\
\hline
Coeff. & CMS exp. & estimated & CMS exp. & estimated &  CMS exp. & estimated \\
\hline
$\fmzero/\Lambda^4$ & [-3.7,3.8] & [-3.9,3.7] & [-7.6,7.6] & input & [-1.0,1.0] & [-1.0,1.0] \\ 
$\fmone/\Lambda^4$ & [-5.4,5.8] & input & [-11,11] & [-11,11] & [-3.0,3.0] & [-3.1,3.1] \\ 
$\fmtwo/\Lambda^4$ & / & / & -- & [-13,13] & -- & [-1.5,1.5] \\ 
$\fmthree/\Lambda^4$ & / & / & -- & [-19,19] & -- & [-5.5,5.5] \\ 
$\fmfour/\Lambda^4$ & / & / & -- & [-5.9,5.9] & -- & [-3.1,3.1] \\ 
$\fmfive/\Lambda^4$ & / & / & -- & [-8.3,8,3] & -- & [-4.5,4.5] \\ 
$\fmseven/\Lambda^4$ & [-8.3,8.1] & [-8.5,8.0] & [-14,14] & [-14,14] & [-5.1,5.1] & input \\ 
\hline
$\fszero/\Lambda^4$ & [-6.0,6.2] & [-6.1,6.2] & [-24,24] & [-25,26] & [-4.2,4.2] & [-6.7,6.8] \\ 
$\fsone/\Lambda^4$ & [-18,19] & [-18,19] & [-38,39] & [-38,39] & [-5.2,5.2] & [-8.3,8.4] \\ 
$\fstwo/\Lambda^4$ & -- & [-18,19] & -- & [-25,26] & -- & [-8.4,8.5] \\ 
\hline
  \end{tabular}
  \caption{Validation for the VBS CMS analyses. All entries represent 95\% CL lower and upper limits in units of TeV$^{-4}$.
  Input limits to the validation procedures, chosen randomly, are highlighted in the table. Bars represent processes for which there is no sensitivity to the corresponding operator, while dashes represent occurrences of an experimental result not quoted in the CMS articles.
  }
  \label{tab:validate} 
  }
\end{table}

The results of the validation phase are presented in Table~\ref{tab:validate}. We first remark that CMS
published results are rather incomplete for both the leptonic
W$^\pm$Z and semileptonic WV analyses, where several
operators for which sensitivity is significant were
not examined. As anticipated, this sensitivity
is in fact comparable to or better than those quoted
in V$\gamma$ VBS analyses~\cite{CMS:2020ioi,CMS:2020ypo}, which therefore we do not
try to reproduce here.

Secondly, we observe that validation is everywhere 
successful except for the S0-2 operators in the
WV semileptonic analysis where validation results are about 60\% more pessimistic with respect to the published values. Several tests were performed
by changing the relative weight of the processes
(especially of W$^+$W$^-$ which has a larger 
background in the experimental analysis) but in none
of them closure was found. We notice, however, that
in the CMS paper details of the simulation are not reported, in particular it is not mentioned whether the latter was performed including b jets or not.
We point out that, when including b jets, other processes
are generated by the syntax 3 in Table~\ref{tab:processes} which are not VBS processes, most notably
electroweak top production via ${\text{q}\mathrm{\overline{q}}\rightarrow \gamma^* \rightarrow\text{t}\mathrm{\overline{t}} \rightarrow \text{W}^+ \text{W}^-\mathrm{b}\mathrm{\overline{b}}}$, and enhance 
the cross section by a factor of $\sim$ 4. While these processes
can be suppressed experimentally via b-jet vetoes, 
there can be a double-counting effect which
artificially leads to lower expected (and observed) limits
if t$\mathrm{\overline{t}}$ backgrounds are 
simulated separately and subtracted.
In conclusion, we do not consider our validation successful 
in this case and cautiously use CMS published limits for scalar operators.

We point out that, as mentioned in item 5 of the above list, the validation procedure has been repeated using different operators as input, and all limits obtained upon input variation have been verified to be essentially identical to those quoted in Table~\ref{tab:validate}. On one hand this confirms the robustness of the whole validation procedure, and on the other hand it allows us to quote limits for the input operators of Table~\ref{tab:validate} to be 
$\fmone/\Lambda^4 \in [-5.3,5.8]$ (VBS W$^\pm$W$^\pm \rightarrow 2\ell 2\nu$), $\fmzero/\Lambda^4 \in [-7.7,7.6]$ (VBS W$^\pm$Z $\rightarrow 3\ell\nu$), and $\fmseven/\Lambda^4 \in [-4.9,4.9]$ (semileptonic VBS), compatibly with the CMS findings.

\section{Implementation of unitarity regularisation in VBS}
\label{sec:unit}

In order to consider unitarity bounds in the
limits, we adopt a different implementation of the clipping
method used in the CMS analysis~\cite{CMS:2020gfh},
following a similar approach as in Refs.~\cite{Kalinowski:2018oxd,Kozow:2019txg,Chaudhary:2019aim}. In our
approach, we evaluate $\mysigma[m_{\min}, m_{\max}]$ for several values of  $m_{\max}$, varying between
$m_{\min}+100$ GeV and the maximum kinematically allowed mass.
In an experimental analysis, this approach would be equivalent to not taking into consideration data or simulated events above $m_{\max}$ in the measurement.
For each of these values of \mysigma, the procedure
used for validation and described in Sec.~\ref{sec:vali} is followed to obtain $m_{\max}$-dependent limits on operator
coefficients. Since only part of the experimental data
fall into the selected $[m_{\min}, m_{\max}]$ invariant-mass intervals, the
95\% CL exclusion limits on the cross section determined
at step 2 are 
rescaled in each test, considering that 
measurement uncertainties at high mass are statistically dominated,
and therefore poissonian errors can be assumed.


\begin{figure}[htbp]
  \centering 
 \includegraphics[width=.46\textwidth]{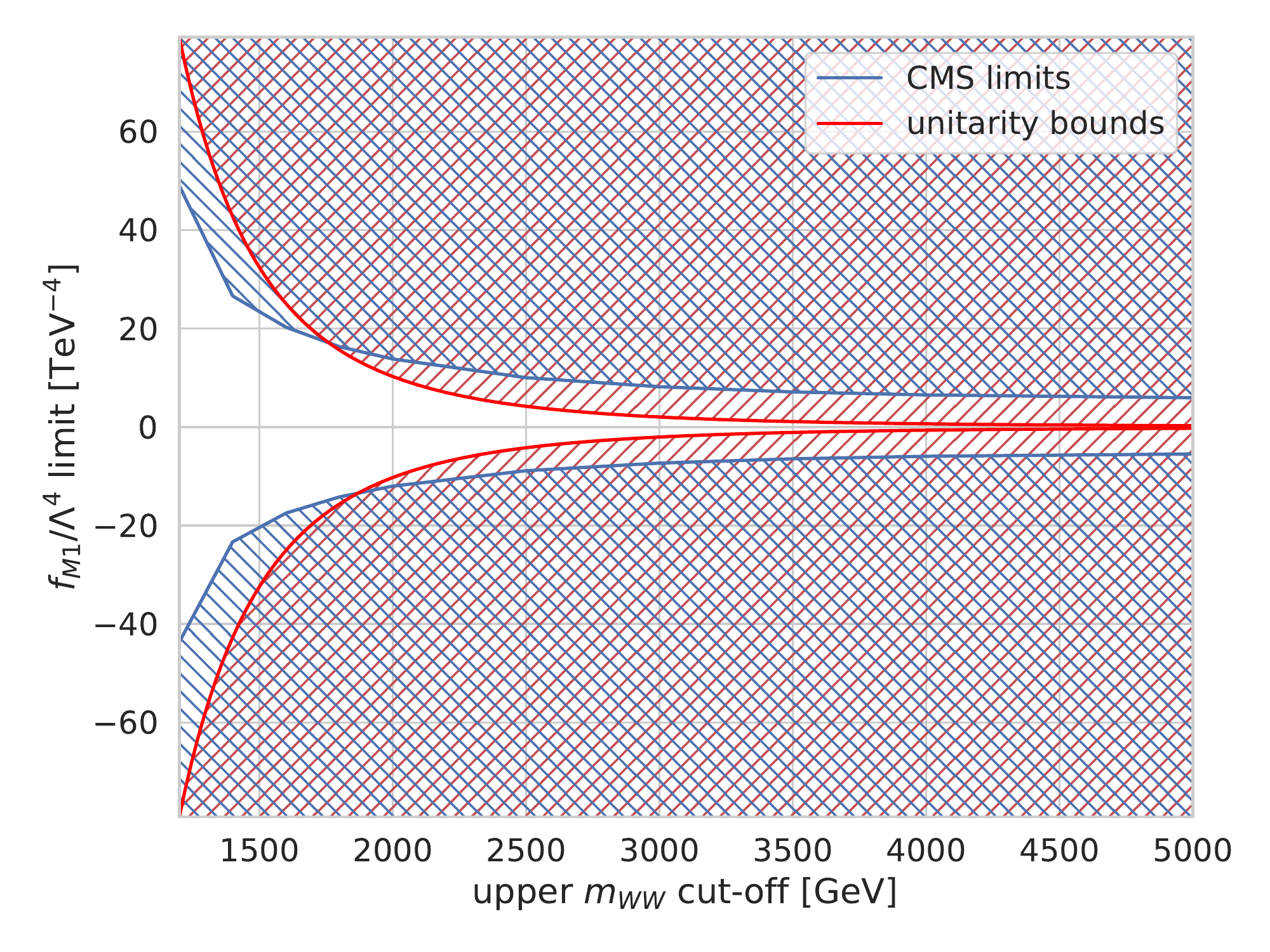}
 \includegraphics[width=.46\textwidth]{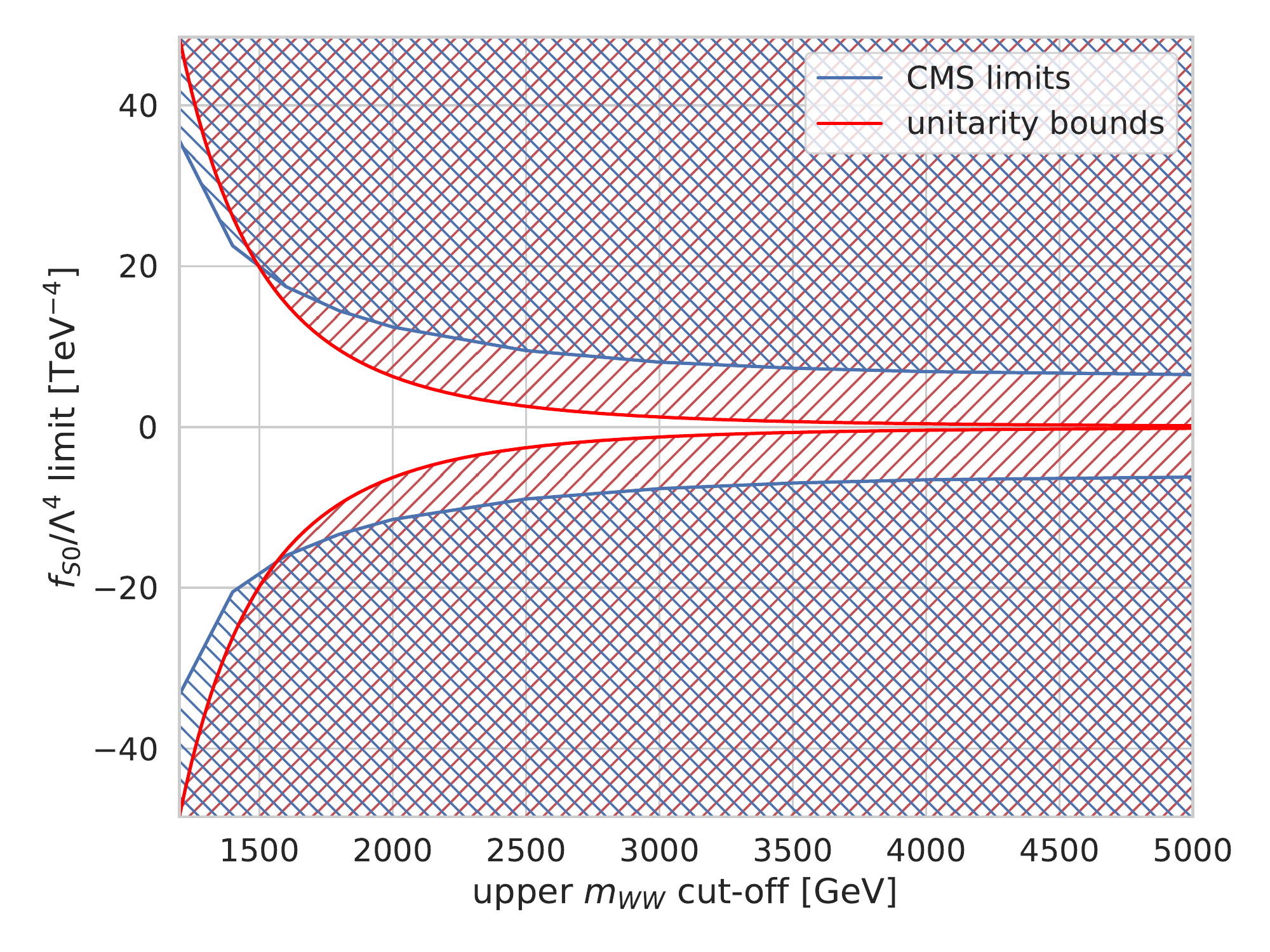} \\
  \includegraphics[width=.46\textwidth]{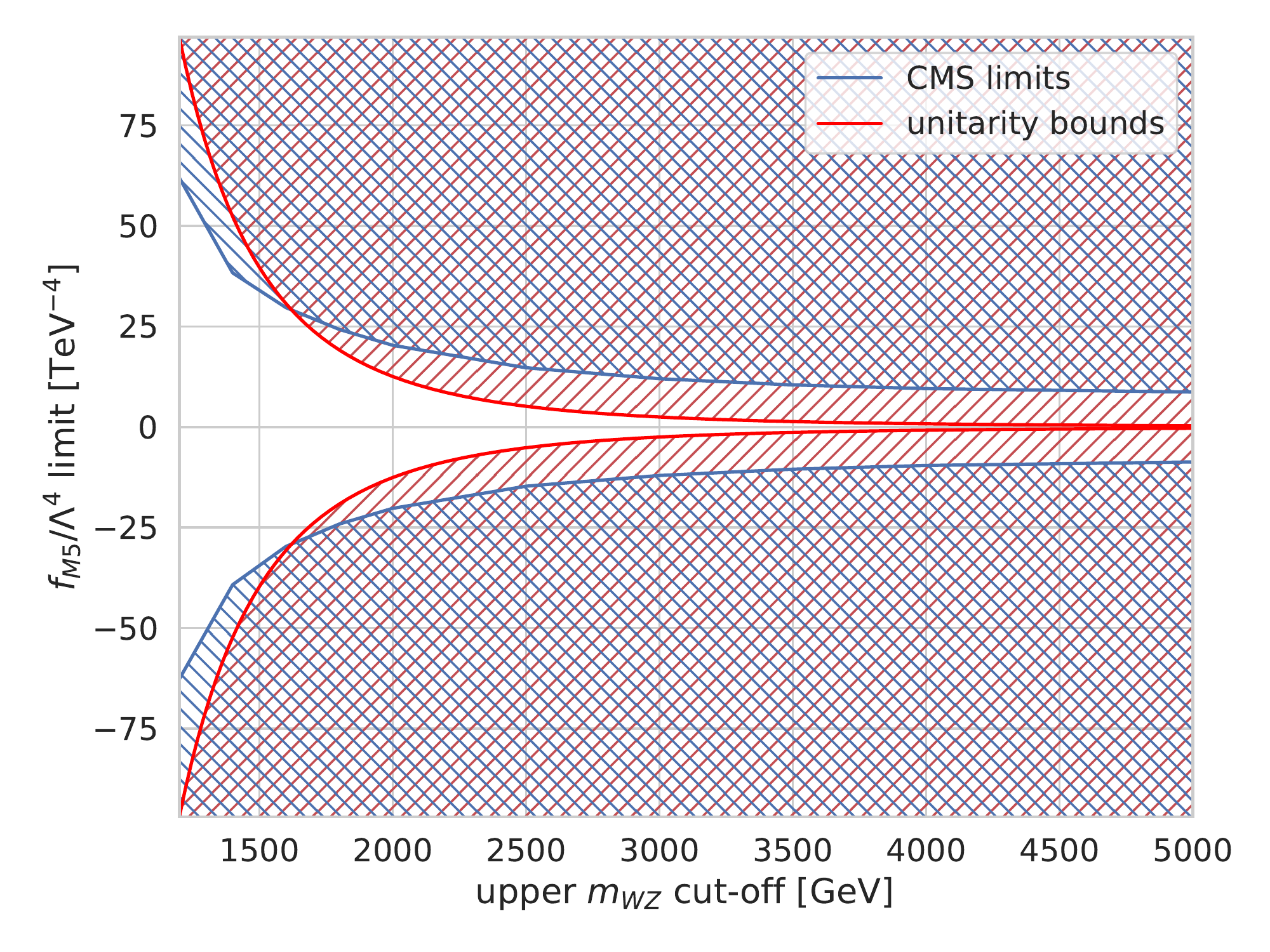}
  \includegraphics[width=.46\textwidth]{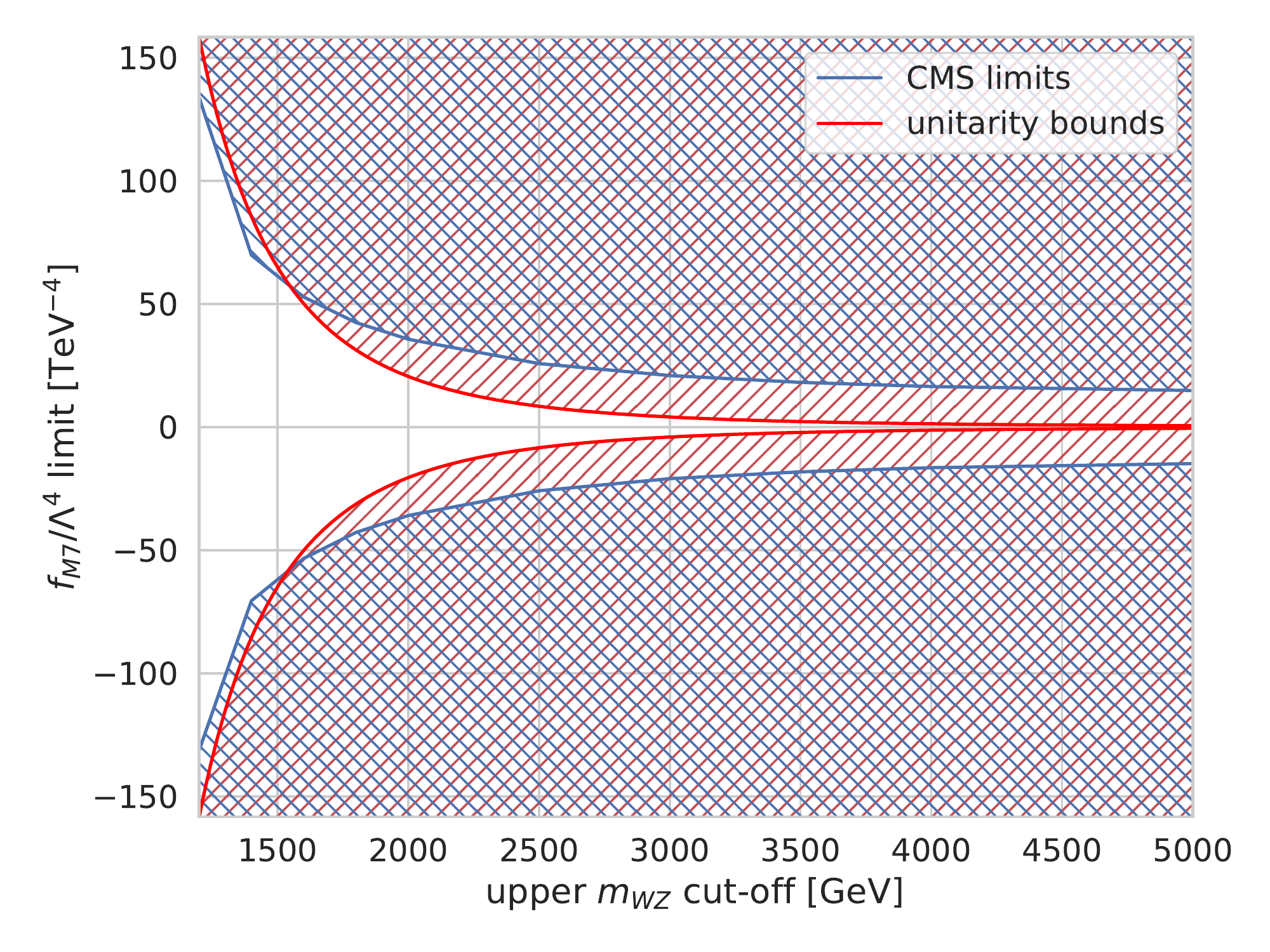} \\
  \includegraphics[width=.46\textwidth]{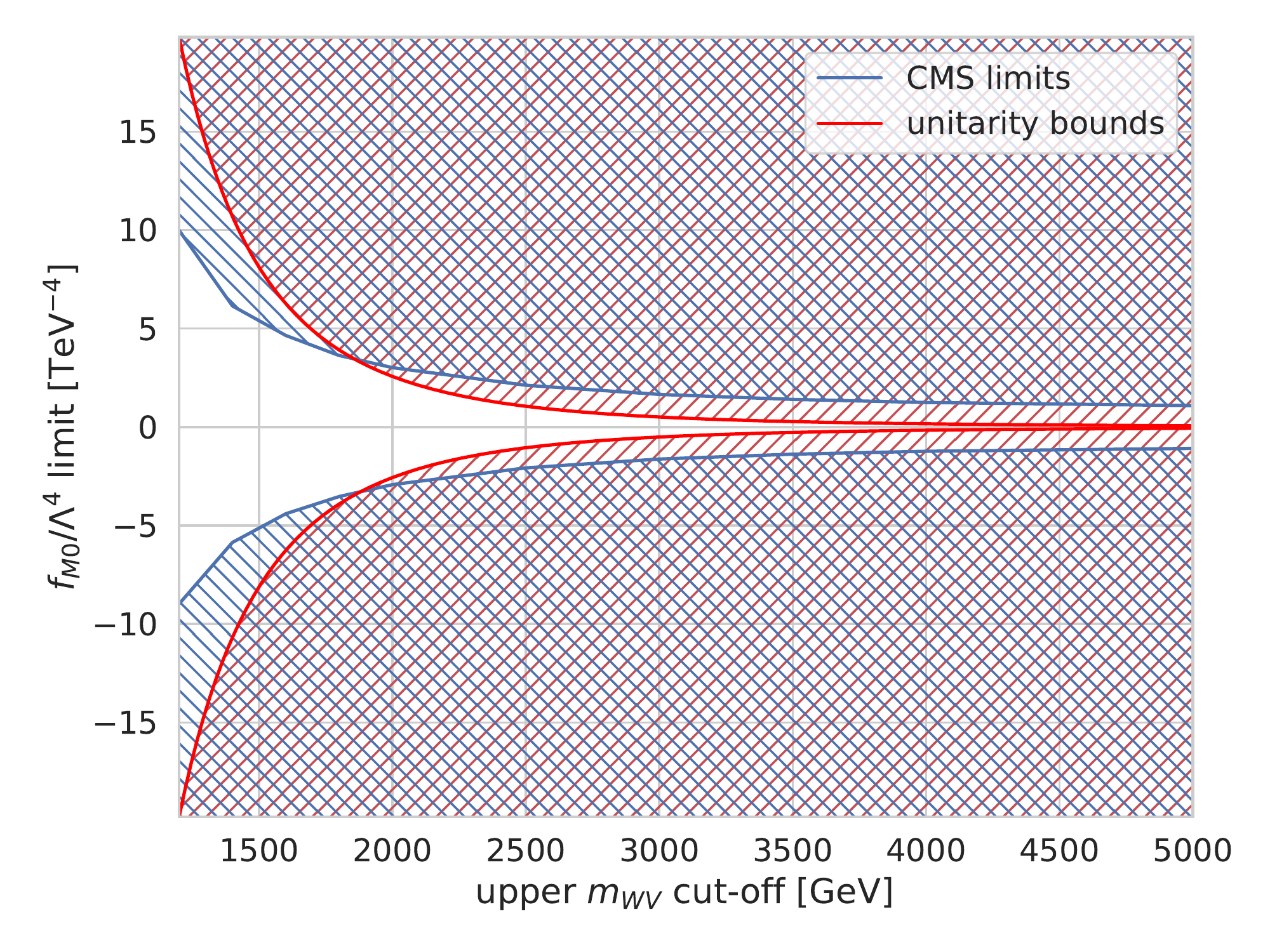} 
  \includegraphics[width=.46\textwidth]{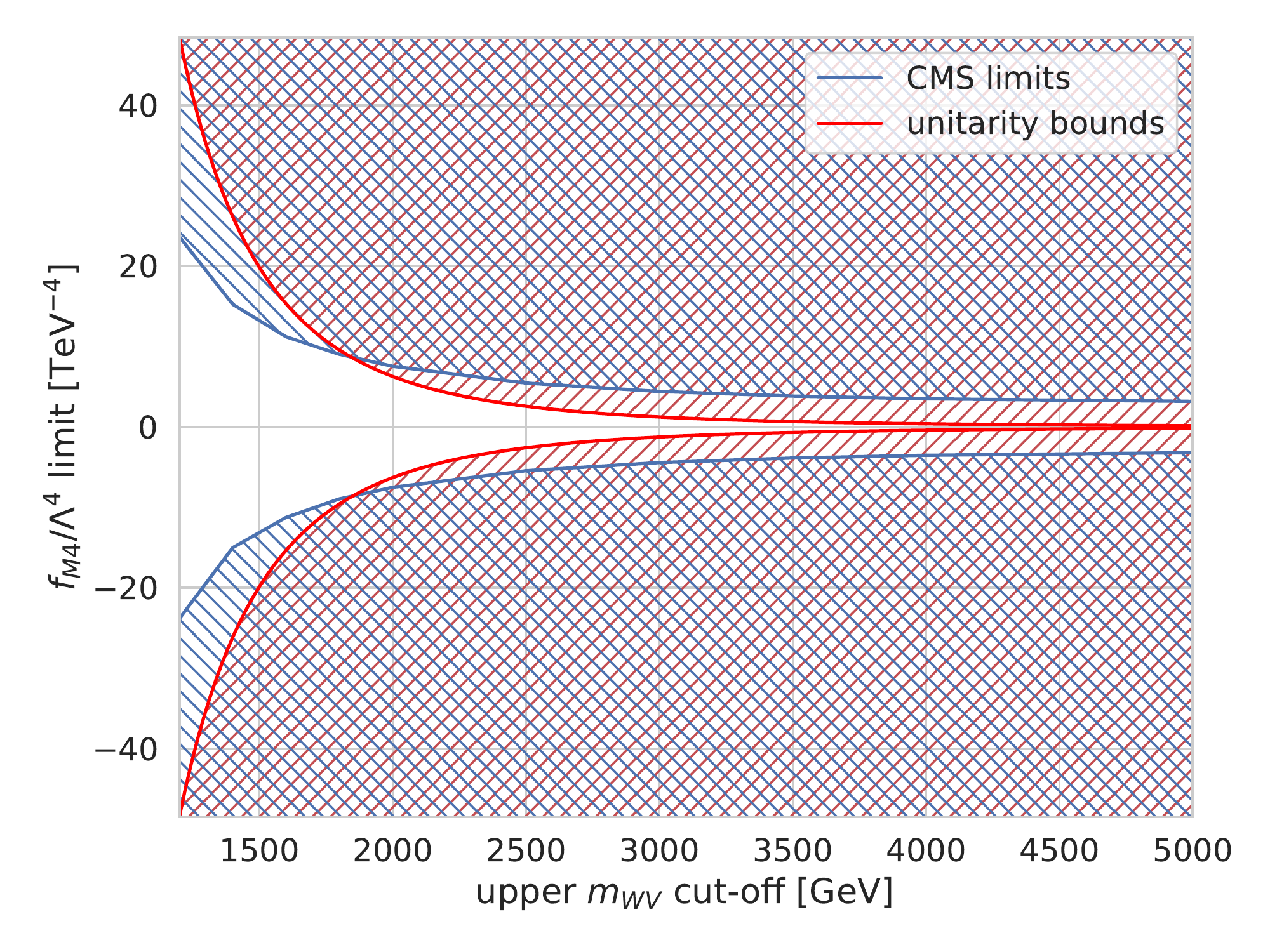}
   \caption{Estimated limits (blue lines) on the Wilson coefficients for a few processes and operators as a function of the upper invariant mass cut-off $m_{\max}$. Superimposed are the unitarity bounds derived from~\cite{Almeida:2020ylr} (red lines). The 
intersection of the experimental and the unitarity-bound curves represents the best limits which can unambiguously be interpreted in an EFT expansion
without violating unitarity. Top row: VBS W$^\pm$W$^\pm \rightarrow 2 \ell 2 \nu$ process, \fmone\ (left), \fszero\ (right).  Middle row: VBS W$^\pm$Z$\rightarrow 3 \ell \nu$ process, \fmfive\ (left), \fmseven\ (right). Bottom row: VBS W$^\pm$V semileptonic process, \fmzero\ (left), \fmfour\ (right).\label{fig:limuni}
}
\end{figure}

\begin{table}[t] {
  \centering
 \begin{tabular}{|c|c|c|c|}
\hline
Coeff. & VBS W$^\pm$W$^\pm$ & VBS W$^\pm$Z & VBS W$^\pm$V semilep. \\
\hline
$\fmzero/\Lambda^4$ & / & / & [-3.3,3.5] \\ 
$\fmone/\Lambda^4$ & [-13,17] & [-67,71] & [-7.4,7.6] \\ 
$\fmtwo/\Lambda^4$ & / & / & [-9.1,9.0] \\ 
$\fmthree/\Lambda^4$ & / & / & [-32,30] \\ 
$\fmfour/\Lambda^4$ & / & [-36,36] & [-8.6,8.7] \\ 
$\fmfive/\Lambda^4$ & / & [-29,29] & [-10,10] \\ 
$\fmseven/\Lambda^4$ & [-21,18] & [-59,57] & [-11,11] \\ 
\hline
$\fszero/\Lambda^4$ & [-17,20] & / & [-8.5,9.5] \\ 
$\fsone/\Lambda^4$ & / & / &  / \\ 
$\fstwo/\Lambda^4$ & / & [-25,26] & [-21,25] \\ 
\hline
  \end{tabular}
  \caption{Results for the VBS CMS analyses when applying unitarity constraints. All entries represent 95\% CL lower and upper limits in units of TeV$^{-4}$.
  Bars represent processes for which there is no sensitivity to the corresponding operator, or cases where the theoretical unitarity bound is more stringent than the experimental one for all $m_{\max}$ cut-off values.
  }
  \label{tab:limuni} 
  }
\end{table}

Fig.~\ref{fig:limuni} shows examples of the $m_{\max}$-dependent limits for specific processes and operators. For large $m_{\max}$ the same
experimental limits shown in Table~\ref{tab:validate} are recovered, while 
the limits become less stringent at smaller $m_{\max}$ 
because of the reduced data statistics as well as of the
weaker dependence of \mysigma\ on the operator coefficient
in the restricted intervals.
In Fig.~\ref{fig:limuni} the unitarity bounds derived
from~\cite{Almeida:2020ylr} are also shown. The 
intersection of the experimental and the unitarity-bound curves represents the maximum
invariant mass that can be used to set experimental
limits which can unambiguously be interpreted in an EFT expansion
without violating unitarity. Hence, the corresponding
limit on $f_X/\Lambda^4$ is the best attainable limit
which can be set using a specific amount of data for
the process under consideration. 
We notice that in all cases these
limits are significantly less stringent than the ones
obtained by neglecting unitarity effects; in some
specific cases, not shown, the two curves do not cross at all, meaning
that the available data are not sufficient yet to 
set more stringent limits than those imposed by
unitarity alone.

In Table~\ref{tab:limuni} we present the best limits
obtained with the outlined procedure for each of the considered
processes and operators. Corresponding CMS limits
using the modified clipping method are quoted in Ref.~\cite{CMS:2020gfh} for
the fully-leptonic results only.
Considering the substantial difference in the unitarity regularisation procedure and in the source of unitarity bounds (which is the VBFNLO program \cite{Arnold:2008rz} in the case of Ref.~\cite{CMS:2020gfh}), it is not surprising that our results display discrepancies with respect to the limits quoted in \cite{CMS:2020gfh}, although we note that the constraint-reducing factors when imposing unitarity
are of similar magnitude in the two analyses.
In particular, \cite{CMS:2020gfh} quotes some very loose constraints on specific operators, which cannot be investigated with our choice
of $m_{\mathrm{min}}$, as they are weaker than the unitarity bounds. These are anyway of modest interest, as coefficient values in such loose ranges would produce too large EFT corrections compared to the SM. 

\section{Results with VBF-HH}
\label{sec:vbfhh}

\begin{figure}[htbp]
  \centering 
  \includegraphics[width=.49\textwidth]{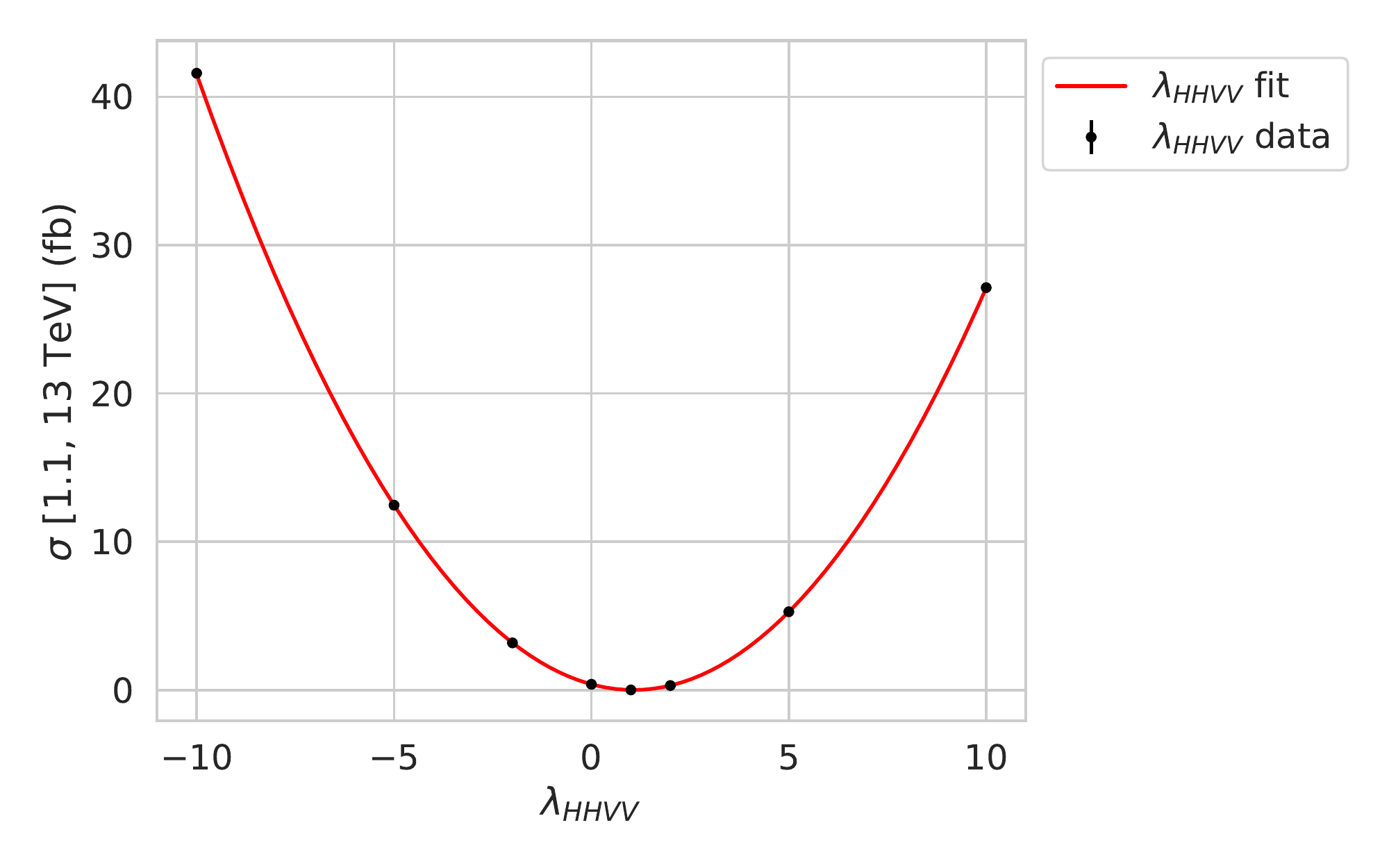}
  \includegraphics[width=.49\textwidth]{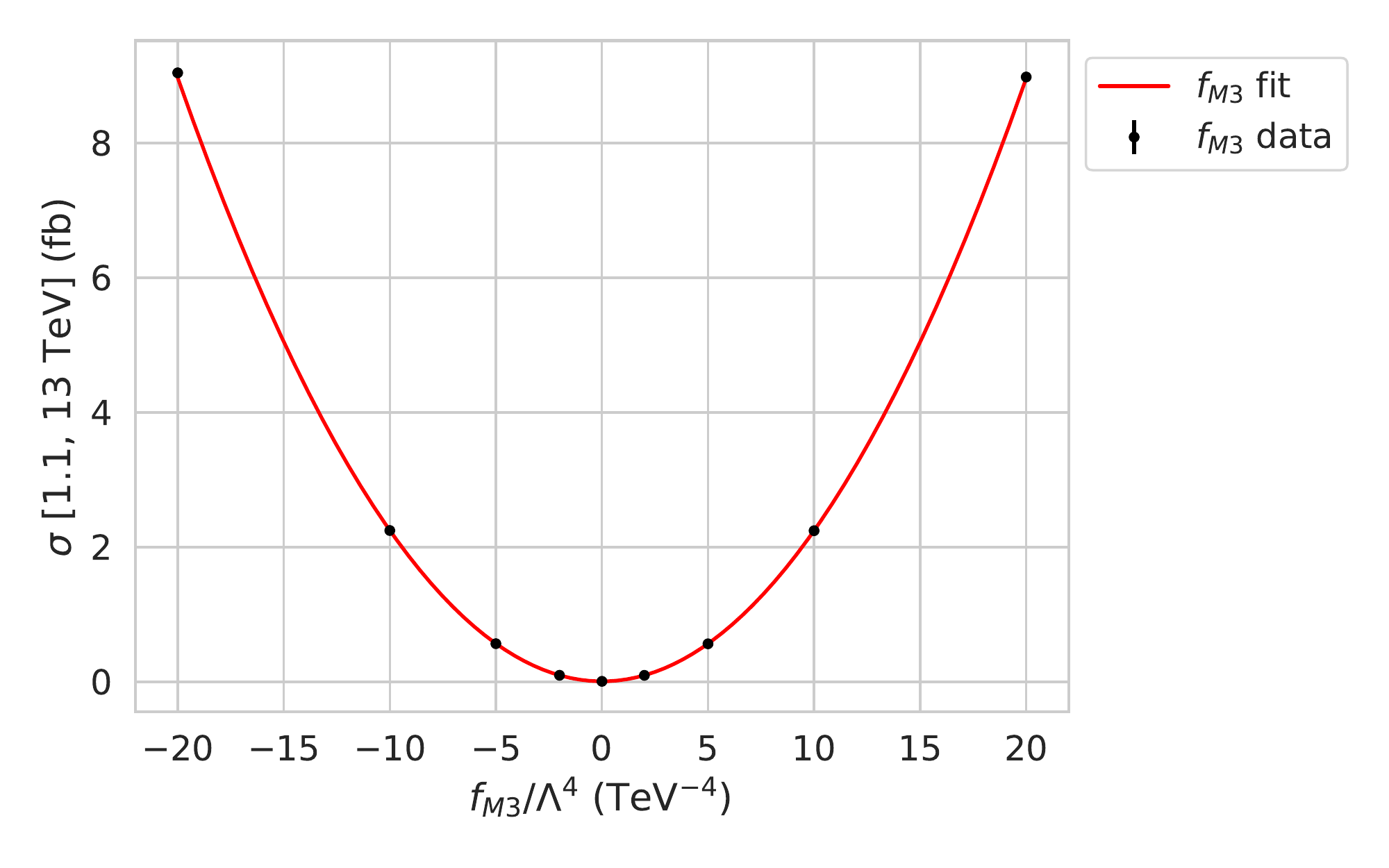} 
   \caption{Quadratic fits of \mysigma[1.1 TeV, 13 TeV] for the VBF-HH process varying
   \kappav\ (left) or \fmthree\ (right).} \label{fig:fitsvbf}  
\end{figure}

As mentioned before, experimental results for 
VBF-HH are given in terms of an effective
VVHH vertex factor, \kappav, rather than as
limits on EFT Wilson coefficients. 
We employ an analogous procedure as the one used 
for the validation, namely:
\begin{enumerate}
\setlength\itemsep{0em} 
    \item The most stringent published 95\% CL exclusion limit on \kappav\ is considered. 
    \item We exploit the VBF-HH
simulation as a function of \kappav\ to impose an exclusion limit on the corresponding parabola, yielding a
    95\% CL exclusion limit\footnote{Translating \kappav\ limits into 
\mysigma\ exclusions could in principle yield
two values, one for the upper one for the lower one,
but these differ by less than 5\%, confirming
the robustness of the simulation.} on \mysigma.
    \item Upon applying such a \mysigma-based exclusion on the parabolae corresponding to all EFT operators, exclusion limits on the corresponding coefficients are
    determined.
    \item As a validation test, by repeating the procedure using one of the $f_X$ coefficients as input, we check that the CMS limits on \kappav\ are reproduced. 
\end{enumerate}
Figure~\ref{fig:fitsvbf} presents \mysigma\ fits
as a function of \kappav\ and of one EFT
Wilson coefficient. In Table~\ref{tab:limvbf}
we present the related upper limits without
unitarity regularisation, compared to the
best limits found using the VBS analysis in Section~\ref{sec:vali}. In spite of the
much smaller yields, we find that, already
with the available LHC data, VBF-HH estimated
limits supersede those obtained with VBS for
\fmzero, \fmtwo\ and \fmthree, and are comparable for most of
the M-type operators. In an experimental
interpretation of the results, 
limits on M-type Wilson coefficients could be 
combined to yield more stringent limits than
those obtained by any single analysis.

\begin{table}[t] {
  \centering
  \begin{tabular}{|c|c|c|c|c|}
\hline
& \multicolumn{2}{|c|}{VBS W$^\pm$V semileptonic} & 
\multicolumn{2}{|c|}{VBF HH$\rightarrow$ b$\mathrm{\overline{b}}$b$\mathrm{\overline{b}}$} \\
\hline
Coeff. & no unitarity & w/ unitarity & no unitarity & w/ unitarity \\
\hline
$\fmzero/\Lambda^4$ & [-1.0,1.0] & [-3.3,3.5] & [-0.95,0.95] & [-3.3,3.3] \\ 
$\fmone/\Lambda^4$ & [-3.1,3.1] & [-7.4,7.6]  & [-3.8,3.8] & [-13,14] \\ 
$\fmtwo/\Lambda^4$ & [-1.5,1.5] & [-9.1,9.0] & [-1.3,1.3] & [-7.6,7.3] \\ 
$\fmthree/\Lambda^4$ & [-5.5,5.5] & [-32,30] & [-5.2,5.3] & [-29,30] \\ 
$\fmfour/\Lambda^4$ & [-3.1,3.1] & [-8.6,8.7] & [-4.0,4.0] & [-14,14] \\ 
$\fmfive/\Lambda^4$ & [-4.5,4.5] & [-10,10] & [-7.1,7.1] & [-26,26] \\ 
$\fmseven/\Lambda^4$ & [-5.1,5.1] & [-11,11] & [-7.6,7.6] & [-27,27] \\ 
\hline
$\fszero/\Lambda^4$ & [-4.2,4.2] & [-8.5,9.5] & [-30,29] & / \\ 
$\fsone/\Lambda^4$ & [-5.2,5.2] & / & [-11,10] & / \\ 
$\fstwo/\Lambda^4$ & - & [-21,25] & [-17,16] & / \\ 
\hline
  \end{tabular}
  \caption{Comparison between VBF-HH constraints and those obtained from semileptonic VBS with or without unitarity constraints, with limits extracted from 
  current CMS Run2 analyses. All entries represent 95\% CL lower and upper limits in units of TeV$^{-4}$.
  Bars represent processes for which there is no sensitivity to the corresponding operator, or cases where the theoretical unitarity bound is more stringent than the experimental one for all $m_{\max}$ cut-off values.
  }
  \label{tab:limvbf} 
  }
\end{table}

\begin{figure}[htbp]
  \centering 
 \includegraphics[width=.46\textwidth]{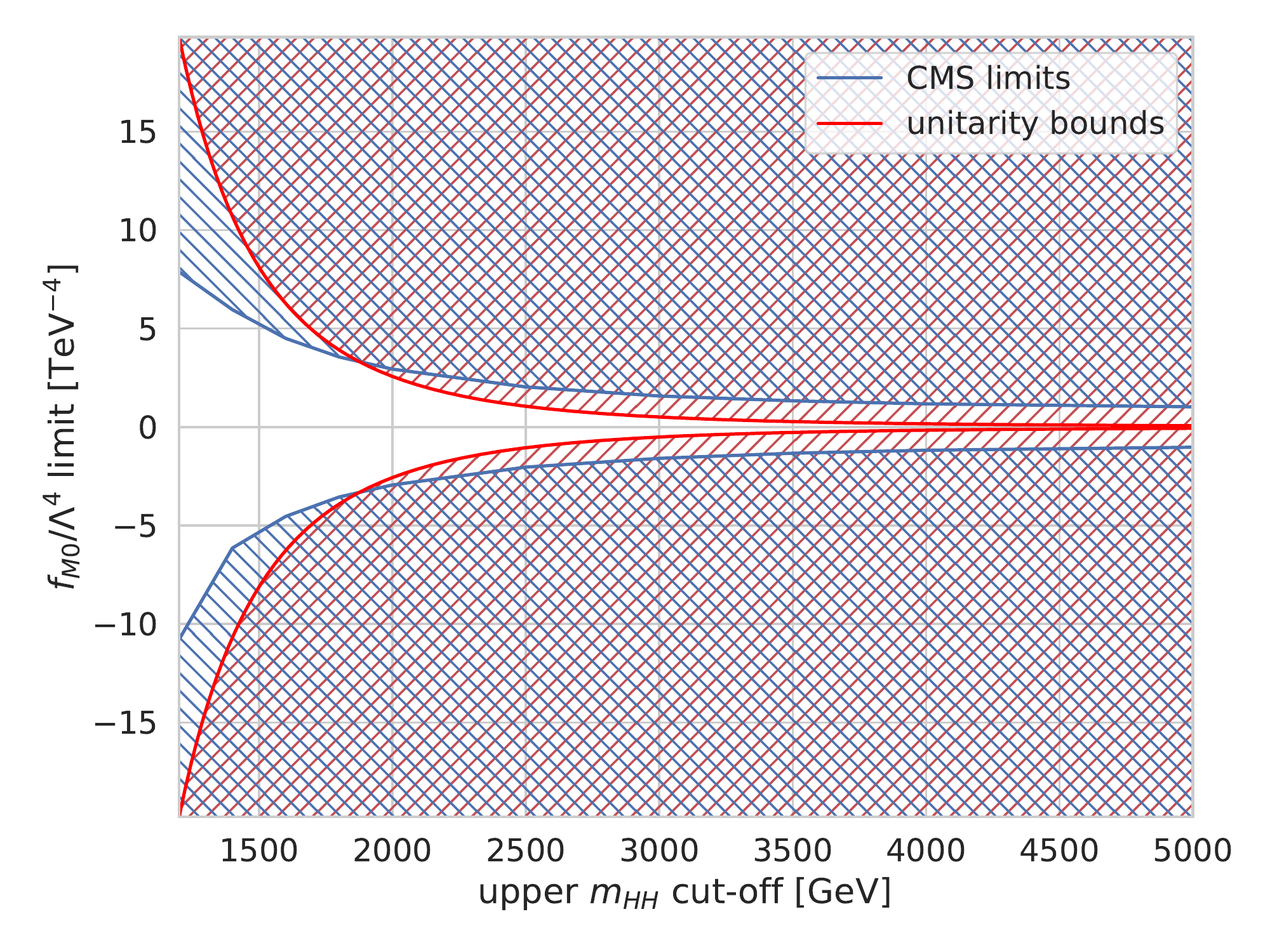}
 \includegraphics[width=.46\textwidth]{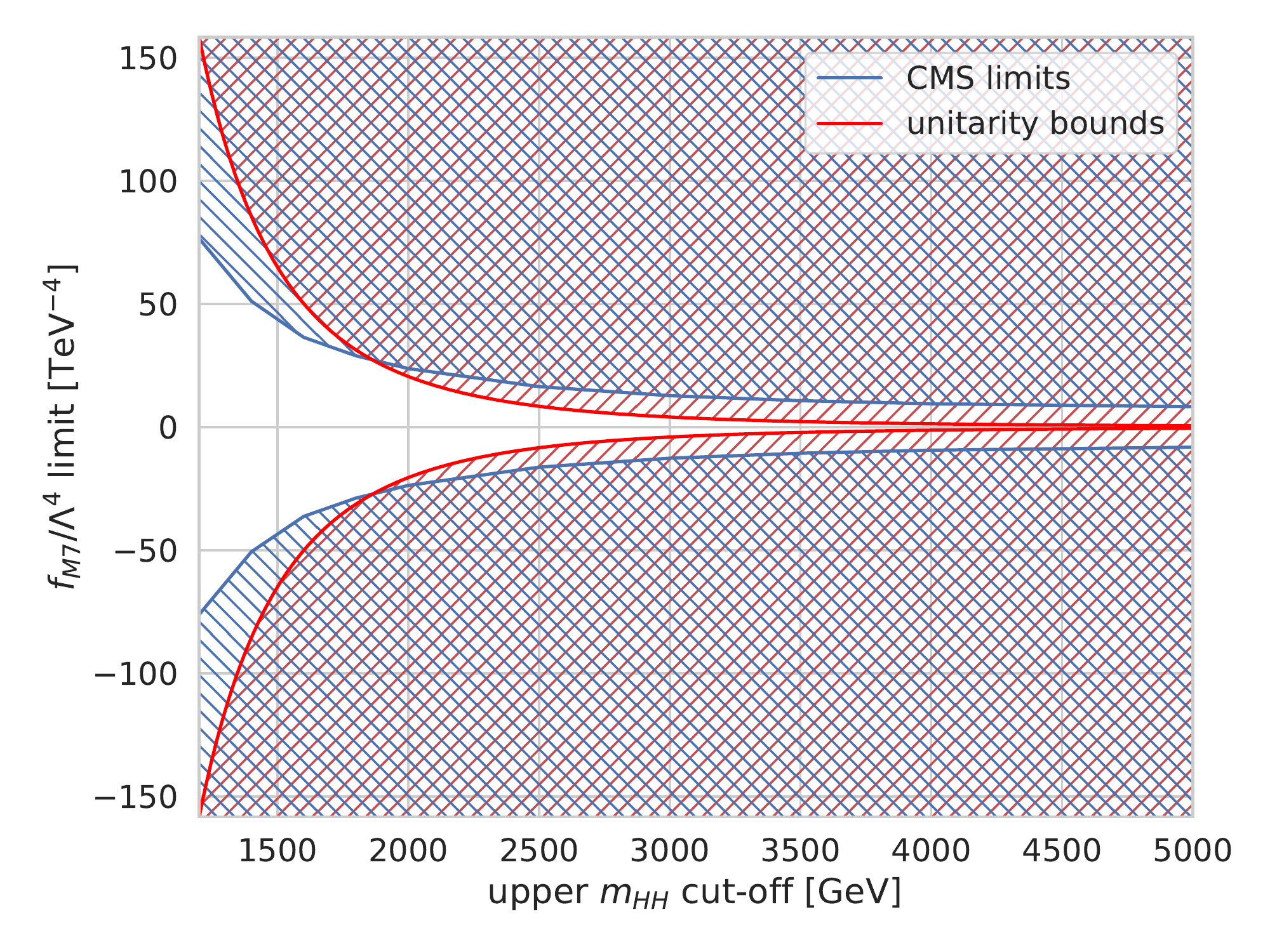} \\
   \caption{Estimated upper and lower limits (blue lines) on the Wilson coefficients in VBF-HH for  \fmzero\ (left) and \fmseven\ (right) as a function of the upper invariant mass cut-off $m_{\max}$. Superimposed are the unitarity bounds derived from~\cite{Almeida:2020ylr} (red lines). The
intersection of the experimental and the unitarity-bound curves represents the best limits which can unambiguously be interpreted in an EFT expansion
without violating unitarity. }\label{fig:limunivbf}  
\end{figure}

An analogous procedure is followed to obtain
limits taking unitarity into account. Figure~\ref{fig:limunivbf} presents 
$m_{\max}$-dependent limits along with
theoretical unitarity bounds. In Table~\ref{tab:limvbf}
we also present the related upper limits with
unitarity regularisation, compared to the
best limits found using the VBS analysis in Section~\ref{sec:unit}. The conclusions of the
previous paragraph remain valid with unitarity as well.

\section{New experimental final states}
\label{sec:zzhzhh}

The \ggzzh\ and ZHH processes, not yet studied in dedicated experimental analyses, are considered in this paper. The purpose is to investigate the potential sensitivity of such processes by performing an exploratory feasibility study. 

\begin{figure}[htbp]
  \centering 
  \includegraphics[width=.76\textwidth]{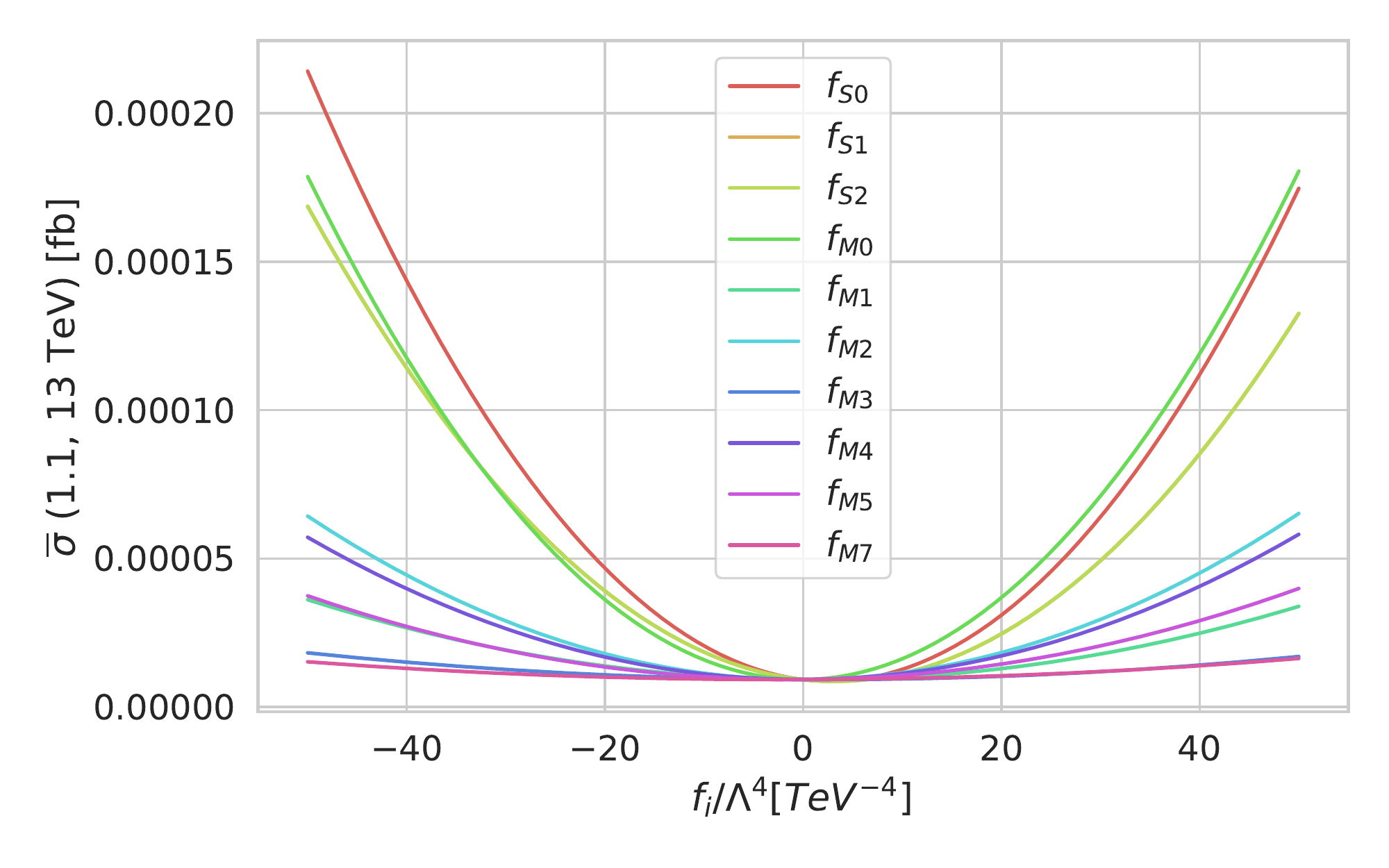}
   \caption{Quadratic fits of \mysigma[1.1 TeV, 13 TeV] for the \ggzzh\ process varying the coefficients of different dimension-8 operators.} 
   \label{fig:fitalloperators}  
\end{figure}

For \ggzzh, we generated samples as described in Section~\ref{sec:simu}, considering EFT modifications of the SM cross section. 
The cross section computed for this process is very small in the SM (see Table~\ref{tab:processes}): in Figure~\ref{fig:fitalloperators} we show the cross section for different values of the EFT coefficients considered. We find that, even for very significant variations ($\pm$40 TeV$^{-4}$) of the EFT Wilson coefficients, the predicted cross section remains extremely small, thus making this process not sensitive enough to be observed at the LHC, not even in the high-luminosity phase. Therefore we refrain from furthering the study of this channel in the present paper.  We stress however that it is possible to simulate this process (as well as other loop-induced or even NLO-accurate processes) in a consistent manner, employing the NLO UFO model we have constructed including the dimension-8 operators from Ref.~\cite{Eboli:2006wa}. 

For the ZZH process, the computed cross section is found to be larger than that of \ggzzh\ in the SM (Table~\ref{tab:processes}), making it worth to estimate the sensitivity that can be obtained from a benchmark analysis of this process. 
In this analysis, we consider the LHC Run2 luminosity and we estimate
the number of detectable events as $N = \mysigma \cdot {\cal L} \cdot \varepsilon \cdot A$,
where $A$ denotes the experimental acceptance for the Z and H decay products and $\varepsilon$
is the corresponding total detection efficiency. 
We compute values of \mysigma\ for the signal and backgrounds described in Section~\ref{sec:simu}, by assuming the Higgs boson decaying into a b$\bar{\text{b}}$ pair, and the Z bosons decaying into leptons (electrons and muons).
We perform Z and H decays using \textsc{Pythia8}~\cite{Bierlich:2022pfr} and define typical LHC
acceptance requirements as: 
\begin{eqnarray}
\pt(\mathrm{b}) &>& 30~\mathrm{GeV}, ~~~~~~~~  |\eta(\mathrm{b})| < 2.5,\nonumber \\
\pt(\mathrm{e},\mu) &>& 20~\mathrm{GeV}, ~~~~~~~~  |\eta(\mathrm{e},\mu)| < 2.4,
\end{eqnarray}
for all decay products, simulating a realistic experimental analysis. The acceptance is computed to be $93\%$ for signal (in the SM, conservatively taken as equal for all EFT scenarios) and $71\%$ for background. The overall efficiency of identification and selection of electrons, muons, and b jets is estimated from experimental papers~\cite{CMS:2021ssj, CMS:2021ugl} to be $70\%$.

From the derived estimates of the signal and background yields, we compute $95\%$ CL upper limits on the cross section of the ZHH process using the Feldman-Cousins approach~\cite{Feldman:1997qc}, and then we use this result to evaluate the limits on EFT operators with a procedure analogous to those presented in the previous sections, taking into account unitarity regularisation.
In Table~\ref{tab:limZHH} we present the values of the limits obtained with such procedure. 
Since no valid limits are found in presence of unitarity with Run2 luminosity, we report as reference the limits obtained without unitarity requirements. We show in the next section that the ZHH process becomes sensitive to EFT variations when considering an increased integrated luminosity.

\begin{table}[t] {
\centering
\begin{tabular}{|c|c|}
\hline
& ZHH $\rightarrow \ell^+\ell^-$b$\mathrm{\overline{b}}$b$\mathrm{\overline{b}}$\\
\hline
Coeff. & no unitarity \\
\hline
$\fmzero/\Lambda^4$ & [-8.4,8.7] \\ 
$\fmone/\Lambda^4$ & [-15,15] \\ 
$\fmtwo/\Lambda^4$ & [-12,12] \\ 
$\fmthree/\Lambda^4$ & [-20,20] \\
$\fmfour/\Lambda^4$ & [-20,21] \\ 
$\fmfive/\Lambda^4$ & [-18,18] \\ 
$\fmseven/\Lambda^4$ & [-29,30] \\
\hline
$\fszero/\Lambda^4$ & [-210,200] \\
$\fsone/\Lambda^4$ & [-350,380] \\ 
$\fstwo/\Lambda^4$ & [-350,380] \\  
\hline 
  \end{tabular}
  \caption{95\% CL lower and upper limits in units of TeV$^{-4}$ for the ZHH process for a CMS Run2 scenario.}
  \label{tab:limZHH} 
  }
\end{table}

\section{Perspectives for HL-LHC}
\label{sec:hllhc}

In spite of the moderate increase in centre-of-mass energy, it is interesting to explore projections for 
the HL-LHC dataset, as the increased
number of events has a non-trivial impact on operator limits when unitarity bounds are accounted for.

In absence of unitarity regularisation, and assuming poissonian scaling in statistically-dominated high-mass regions, 95\% CL exclusion limits on \mysigma\ are expected to improve as ${\cal L}^{-1/2}$, with ${\cal L}$ being the
integrated luminosity. Since 
the dependence of the cross section on the Wilson
coefficients is quadratic, the 
related $f_X/\Lambda^4$ limits scale as ${\cal L}^{-1/4}$, 
yielding an improvement of a factor about 2.2 when passing from the 140 fb$^{-1}$
of the LHC Run2 to the 3000 fb$^{-1}$ expected from the HL-LHC\footnote{We neglect the increase in $\sqrt{s}$ in this estimate.}.

\begin{figure}[htbp]
  \centering 
 \includegraphics[width=.46\textwidth]{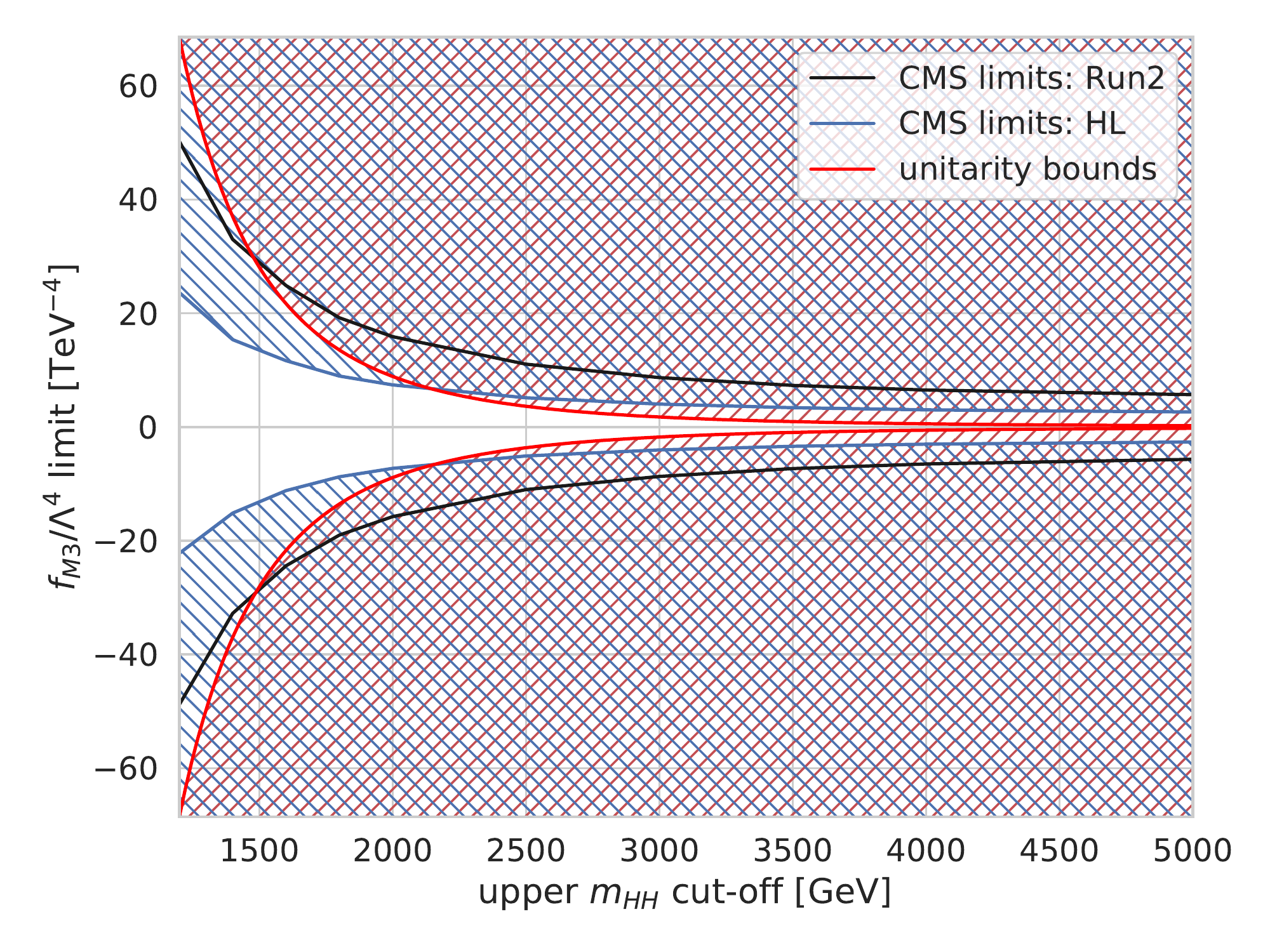}
  \includegraphics[width=.46\textwidth]{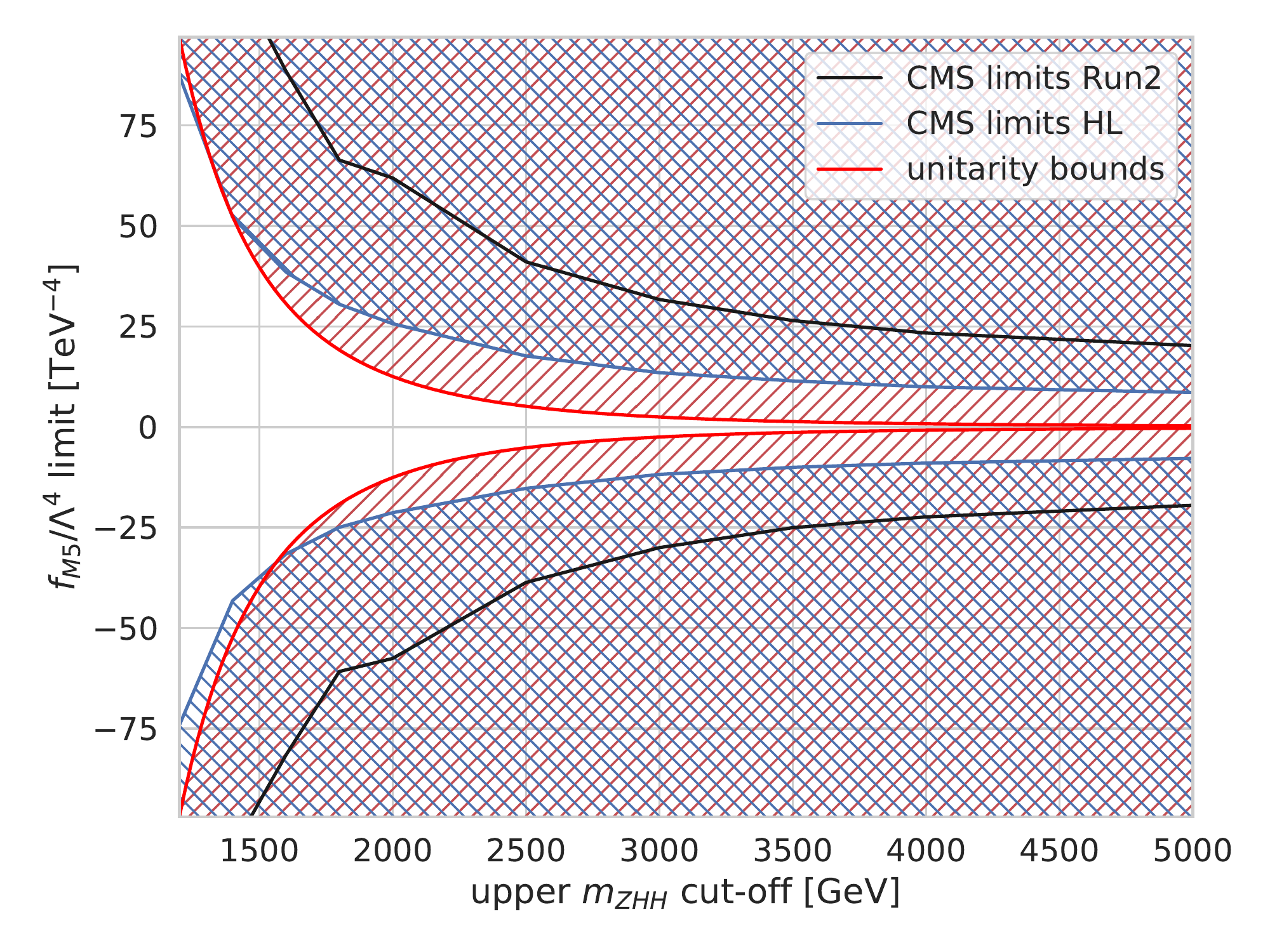}
   \caption{Estimated limits on the \fmthree\  Wilson coefficient in VBF-HH (left) and \fmfive\ in ZHH (right) as a function of the upper invariant mass cut-off $m_{\max}$. Black lines represent the current Run2 scenario (140 fb$^{-1}$) while blue lines represent the HL-LHC scenario (3000 fb$^{-1}$). Superimposed are the unitarity bounds derived from~\cite{Almeida:2020ylr} (red lines). The
intersection of the experimental and the unitarity-bound curves represents the best limits which can unambiguously be interpreted in an EFT expansion
without violating unitarity. While all individual results
scale roughly as ${\cal L}^{-1/4}$, the improvement on the EFT-interpretable limit is a factor of 4.6.}\label{fig:hllhc}  
\end{figure}

In Table~\ref{tab:limhllhc}
we present the expected limits in a HL-LHC scenario, with and without
unitarity regularisation, for VBF-HH compared to the
best limits found using the VBS analysis of Section~\ref{sec:vali}. Limits in absence of unitarity are obtained by simply rescaling the excluded cross sections by the square-root ratio of the integrated luminosities, hence the sensitivity hierarchy in this case is the same as in the Run2 study. 
Conversely, limits in presence of unitarity have a significant gain from an increased dataset since, while 95\% CL limits become tighter, the $m_{\max}$ value for which
the result remains EFT-interpretable correspondingly moves to larger values,
allowing more data to be included in the sensitivity estimate. 
This is due to the fact that a higher integrated luminosity enhances the sensitivity to deviations from the SM that are generated by smaller operator coefficients, resulting in a wider invariant-mass window allowed by unitarity constraints.
This effect would lead to the first physical limit on \fsone, as well as to improvements
on other limits by factors of up to 4-5, as shown for \fmthree\ in Figure~\ref{fig:hllhc}~(left).

\begin{table}[t] {
  \centering
  \begin{tabular}{|c|c|c|c|c|}
\hline
& \multicolumn{2}{|c|}{VBS W$^\pm$V semileptonic} & 
\multicolumn{2}{|c|}{VBF HH$\rightarrow$ b$\mathrm{\overline{b}}$b$\mathrm{\overline{b}}$} \\
\hline
Coeff. & no unitarity & w/ unitarity & no unitarity & w/ unitarity \\
\hline
$\fmzero/\Lambda^4$ & [-0.47,0.47] & [-0.96,1.02] & [-0.43,0.43] & [-0.90,0.87] \\ 
$\fmone/\Lambda^4$ & [-1.5,1.5] & [-2.3,2.4]  & [-1.7,1.7] & [-3.5,3.5] \\ 
$\fmtwo/\Lambda^4$ & [-0.69,0.68] & [-2.1,2.1] & [-0.62,0.61] & [-1.7,1.7] \\ 
$\fmthree/\Lambda^4$ & [-2.5,2.4] & [-6.8,6.3] & [-2.4,2.4] & [-6.5,6.6] \\ 
$\fmfour/\Lambda^4$ & [-1.4,1.4] & [-2.4,2.5] & [-1.8,1.8] & [-3.9,4.0] \\ 
$\fmfive/\Lambda^4$ & [-2.0,2.0] & [-3.0,3.1] & [-3.2,3.2] & [-6.9,7.0] \\ 
$\fmseven/\Lambda^4$ & [-2.4,2.4] & [-3.5,3.5] & [-3.5,3.5] & [-7.1,7.1] \\ 
\hline
$\fszero/\Lambda^4$ & [-1.8,2.0] & [-2.6,3.3] & [-14,13] & / \\ 
$\fsone/\Lambda^4$ & [-2.4,2.4] & [-5.8,6.1] & [-5.1,4.5] & / \\ 
$\fstwo/\Lambda^4$ & [-2.3,2.4] & [-4.8,5.2] & [-8.1,7.1] & / \\ 
\hline
  \end{tabular}
  \caption{Comparison of VBF-HH constraints with those obtained from VBS when applying or not unitarity constraints, in a projection with  full HL-LHC luminosity. All entries represent 95\% CL lower and upper limits in units of TeV$^{-4}$.
  Bars represent processes for which there is no sensitivity to the corresponding operator, or cases where the theoretical unitarity bound is more stringent than the experimental one for all $m_{\max}$ cut-off values.
  }
  \label{tab:limhllhc} 
  }
\end{table}

Table~\ref{tab:limhllhczhh} reports the 95\% CL limits obtained  for the ZHH process in the HL-LHC scenario, with and without unitarity regularisation. 
As anticipated in Section~\ref{sec:zzhzhh}, this process becomes sensitive to dimension-8 operators when increasing the luminosity. 
In this scenario, in fact, it is possible to set limits on some M-type operators even in presence of unitarity bounds, as shown in Figure~\ref{fig:hllhc}~(right).
We note that such limits are not quite competitive, due to the small cross section of the ZHH process with respect to its background process: for future analyses, it will be crucial to develop strategies to enhance the signal and reduce the number of background events. As a simple test, by halving the b-jet pair invariant-mass window for the background, described in Section~\ref{sec:simu}, we observe an improvement of about 20\% on the limits both with and without unitarity regularisation, stemming from the decrease in the background contribution.

\begin{table}[t] {
\centering
\begin{tabular}{|c|c|c|}
\hline
& \multicolumn{2}{|c|}{ZHH$\rightarrow\ell^+\ell^-$ b$\mathrm{\overline{b}}$b$\mathrm{\overline{b}}$} \\
\hline
Coeff. & no unitarity & w/ unitarity\\
\hline
$\fmzero/\Lambda^4$ & [-3.4,3.7] & / \\ 
$\fmone/\Lambda^4$ & [-6.4,5.9] & [-66,31]\\ 
$\fmtwo/\Lambda^4$ & [-4.7,4.8] & / \\ 
$\fmthree/\Lambda^4$ & [-8.4,8.2] & / \\ 
$\fmfour/\Lambda^4$ & [-8.2,8.9] & / \\ 
$\fmfive/\Lambda^4$ & [-7.1,7.7] & [-34,52]\\ 
$\fmseven/\Lambda^4$ & [-12,13] & [-91,160]\\ 
\hline
$\fszero/\Lambda^4$ & [-90,83] & / \\ 
$\fsone/\Lambda^4$ & [-140,160] & / \\ 
$\fstwo/\Lambda^4$ & [-140,160] & / \\ 
\hline
  \end{tabular}
  \caption{Constraints obtained with ZHH when applying or not unitarity constraints, in a projection with  full HL-LHC luminosity. All entries represent 95\% CL lower and upper limits in units of TeV$^{-4}$.
  Bars represent processes for which there is no sensitivity to the corresponding operator, or cases where the theoretical unitarity bound is more stringent than the experimental one for all $m_{\max}$ cut-off values.
  \label{tab:limhllhczhh} 
  }
  }
\end{table}

\section{Summary}
\label{sec:summ}
In this article we have studied the sensitivity to BSM effects of the interaction between two electroweak gauge bosons (V) and two Higgs bosons (H).
To this aim, we have focused on hadronic reactions that feature the VVHH vertex at lowest order, in particular VBF Higgs-pair production, ZHH associated production, and loop-induced gluon-fusion ZZH production. We have considered modifications to these processes induced by dimension-8 EFT operators affecting the VVHH interaction, and set up a simplified framework to cast stringent bounds on the Wilson coefficients of these operators using such reactions.

After validating our framework against published experimental limits, we have performed a systematic comparison of our results with current bounds from the CMS collaboration, based on the VBS signature. This has shown that, as far as  dimension-8 VVHH-modifiers are concerned, the VBF-HH channel is able to yield limits which are comparable to, or even more stringent than, those stemming from VBS, thus proving viable in view of a full-fledged experimental analysis, as well as of an experimental combination with VBS limits. The ZHH and \ggzzh\ channel have been instead shown to have more limited constraining power on the same operators with Run2 luminosity.

We have then considered the effect of unitarity constraints on the extraction of exclusion limits in an EFT-based analysis. As this subject is typically treated in a non-systematic manner in current extractions, we have employed a dedicated prescription to consistently take unitarity bounds into account. We find that, as expected, unitarity constraints systematically weaken the experimental limits on EFT Wilson coefficients, including the ones set by current CMS analyses of the VBS channel. However, even in presence of unitarity, VBF-HH limits are equally competitive with VBS-based ones.

We have finally investigated the perspectives of the exclusion limits for the high-luminosity phase of the LHC, showing that limits in presence of unitarity can improve at the HL-LHC by a factor of up to 5 with respect to Run2, which is significantly more than what is expected from pure luminosity-scaling considerations. In this scenario, the ZHH final states also can contribute in a combined exclusion of some of the relevant EFT coefficients.

\acknowledgments

We are grateful to Ilaria Brivio and to Giampiero Passarino for useful discussions and a careful reading of the manuscript.

AC has received funding from the European Union’s Horizon 2020 research and innovation programme under the Marie Skłodowska-Curie grant agreement No 899987. 
RC
acknowledges support from the Italian and Serbian Ministries for Foreign Affairs through the
Researcher Mobility Program RS19MO06.
PT has been partially supported by the Italian Ministry of University
and Research (MIUR) through grant PRIN 20172LNEEZ and by Compagnia di
San Paolo through grant TORP\_S1921\_EX-POST\_21\_01.
MZ is supported by the "Programma per Giovani Ricercatori Rita Levi Montalcini" granted by the Italian Ministero dell’Universit\`a e della Ricerca (MUR).

\bibliographystyle{JHEP}
\bibliography{ZZHH}

\providecommand{\href}[2]{#2}\begingroup\raggedright\begin{thebibliography}{10}

\bibitem{Belanger:1992qh}
G.~Belanger and F.~Boudjema, \emph{{Probing quartic couplings of weak bosons
  through three vectors production at a 500-GeV NLC}},
  \href{https://doi.org/10.1016/0370-2693(92)91978-I}{\emph{Phys. Lett. B}
  {\bfseries 288} (1992) 201}.

\bibitem{Belanger:1992qi}
G.~Belanger and F.~Boudjema, \emph{{gamma gamma ---\ensuremath{>} W+ W- and
  gamma gamma ---\ensuremath{>} Z Z as tests of novel quartic couplings}},
  \href{https://doi.org/10.1016/0370-2693(92)91979-J}{\emph{Phys. Lett. B}
  {\bfseries 288} (1992) 210}.

\bibitem{Eboli:2003nq}
O.J.P.~Eboli, M.C.~Gonzalez-Garcia and S.M.~Lietti, \emph{{Bosonic quartic
  couplings at CERN LHC}},
  \href{https://doi.org/10.1103/PhysRevD.69.095005}{\emph{Phys. Rev. D}
  {\bfseries 69} (2004) 095005}
  [\href{https://arxiv.org/abs/hep-ph/0310141}{{\ttfamily hep-ph/0310141}}].

\bibitem{Degrande:2012wf}
C.~Degrande, N.~Greiner, W.~Kilian, O.~Mattelaer, H.~Mebane, T.~Stelzer et~al.,
  \emph{{Effective Field Theory: A Modern Approach to Anomalous Couplings}},
  \href{https://doi.org/10.1016/j.aop.2013.04.016}{\emph{Annals Phys.}
  {\bfseries 335} (2013) 21} [\href{https://arxiv.org/abs/1205.4231}{{\ttfamily
  1205.4231}}].

\bibitem{Falkowski:2016cxu}
A.~Falkowski, M.~Gonzalez-Alonso, A.~Greljo, D.~Marzocca and M.~Son,
  \emph{{Anomalous Triple Gauge Couplings in the Effective Field Theory
  Approach at the LHC}},
  \href{https://doi.org/10.1007/JHEP02(2017)115}{\emph{JHEP} {\bfseries 02}
  (2017) 115} [\href{https://arxiv.org/abs/1609.06312}{{\ttfamily
  1609.06312}}].

\bibitem{Nordstrom:2018ceg}
K.~Nordstr\"om and A.~Papaefstathiou, \emph{{$VHH$ production at the
  High-Luminosity LHC}},
  \href{https://doi.org/10.1140/epjp/i2019-12614-2}{\emph{Eur. Phys. J. Plus}
  {\bfseries 134} (2019) 288}
  [\href{https://arxiv.org/abs/1807.01571}{{\ttfamily 1807.01571}}].

\bibitem{Arganda:2018ftn}
E.~Arganda, C.~Garcia-Garcia and M.J.~Herrero, \emph{{Probing the Higgs
  self-coupling through double Higgs production in vector boson scattering at
  the LHC}}, \href{https://doi.org/10.1016/j.nuclphysb.2019.114687}{\emph{Nucl.
  Phys. B} {\bfseries 945} (2019) 114687}
  [\href{https://arxiv.org/abs/1807.09736}{{\ttfamily 1807.09736}}].

\bibitem{Buchmuller:1985jz}
W.~Buchmuller and D.~Wyler, \emph{{Effective Lagrangian Analysis of New
  Interactions and Flavor Conservation}},
  \href{https://doi.org/10.1016/0550-3213(86)90262-2}{\emph{Nucl. Phys. B}
  {\bfseries 268} (1986) 621}.

\bibitem{Grzadkowski:2010es}
B.~Grzadkowski, M.~Iskrzynski, M.~Misiak and J.~Rosiek, \emph{{Dimension-Six
  Terms in the Standard Model Lagrangian}},
  \href{https://doi.org/10.1007/JHEP10(2010)085}{\emph{JHEP} {\bfseries 10}
  (2010) 085} [\href{https://arxiv.org/abs/1008.4884}{{\ttfamily 1008.4884}}].

\bibitem{Hagiwara:1993ck}
K.~Hagiwara, S.~Ishihara, R.~Szalapski and D.~Zeppenfeld, \emph{{Low-energy
  effects of new interactions in the electroweak boson sector}},
  \href{https://doi.org/10.1103/PhysRevD.48.2182}{\emph{Phys. Rev. D}
  {\bfseries 48} (1993) 2182}.

\bibitem{Gonzalez-Garcia:1999ije}
M.C.~Gonzalez-Garcia, \emph{{Anomalous Higgs couplings}},
  \href{https://doi.org/10.1142/S0217751X99001494}{\emph{Int. J. Mod. Phys. A}
  {\bfseries 14} (1999) 3121}
  [\href{https://arxiv.org/abs/hep-ph/9902321}{{\ttfamily hep-ph/9902321}}].

\bibitem{Green:2016trm}
D.R.~Green, P.~Meade and M.-A.~Pleier, \emph{{Multiboson interactions at the
  LHC}}, \href{https://doi.org/10.1103/RevModPhys.89.035008}{\emph{Rev. Mod.
  Phys.} {\bfseries 89} (2017) 035008}
  [\href{https://arxiv.org/abs/1610.07572}{{\ttfamily 1610.07572}}].

\bibitem{CMS:2021icx}
{\scshape CMS} collaboration, \emph{{Measurement of the inclusive and
  differential WZ production cross sections, polarization angles, and triple
  gauge couplings in pp collisions at $\sqrt{s}$ = 13 TeV}},
  \href{https://arxiv.org/abs/2110.11231}{{\ttfamily 2110.11231}}.

\bibitem{CMS:2021foa}
{\scshape CMS} collaboration, \emph{{Measurement of the W$\gamma$ Production
  Cross Section in Proton-Proton Collisions at $\sqrt {s}$=13\,\,TeV and
  Constraints on Effective Field Theory Coefficients}},
  \href{https://doi.org/10.1103/PhysRevLett.126.252002}{\emph{Phys. Rev. Lett.}
  {\bfseries 126} (2021) 252002}
  [\href{https://arxiv.org/abs/2102.02283}{{\ttfamily 2102.02283}}].

\bibitem{CMS:2020mxy}
{\scshape CMS} collaboration, \emph{{W$^+$W$^-$ boson pair production in
  proton-proton collisions at $\sqrt{s} =$ 13 TeV}},
  \href{https://doi.org/10.1103/PhysRevD.102.092001}{\emph{Phys. Rev. D}
  {\bfseries 102} (2020) 092001}
  [\href{https://arxiv.org/abs/2009.00119}{{\ttfamily 2009.00119}}].

\bibitem{CMS:2019ppl}
{\scshape CMS} collaboration, \emph{{Search for anomalous triple gauge
  couplings in WW and WZ production in lepton + jet events in proton-proton
  collisions at $\sqrt{s} =$ 13 TeV}},
  \href{https://doi.org/10.1007/JHEP12(2019)062}{\emph{JHEP} {\bfseries 12}
  (2019) 062} [\href{https://arxiv.org/abs/1907.08354}{{\ttfamily
  1907.08354}}].

\bibitem{ATLAS:2021jgw}
{\scshape ATLAS} collaboration, \emph{{Measurements of $W^+W^-+\ge 1~$jet
  production cross-sections in $pp$ collisions at $\sqrt{s}=13~$TeV with the
  ATLAS detector}}, \href{https://doi.org/10.1007/JHEP06(2021)003}{\emph{JHEP}
  {\bfseries 06} (2021) 003}
  [\href{https://arxiv.org/abs/2103.10319}{{\ttfamily 2103.10319}}].

\bibitem{ATLAS:2021ohb}
{\scshape ATLAS} collaboration, ``{Combined effective field theory
  interpretation of differential cross-sections measurements of $WW$, $WZ$,
  4$\ell$, and $Z$-plus-two-jets production using ATLAS data,
  ATL-PHYS-PUB-2021-022}.'' 2021.

\bibitem{CMS:2019nep}
{\scshape CMS} collaboration, \emph{{Measurement of electroweak production of a
  $\mathrm{W} $ boson in association with two jets in proton\textendash{}proton
  collisions at $\sqrt{s}=13\,\text {Te}\text {V} $}},
  \href{https://doi.org/10.1140/epjc/s10052-019-7585-7}{\emph{Eur. Phys. J. C}
  {\bfseries 80} (2020) 43} [\href{https://arxiv.org/abs/1903.04040}{{\ttfamily
  1903.04040}}].

\bibitem{CMS:2017dmo}
{\scshape CMS} collaboration, \emph{{Electroweak production of two jets in
  association with a Z boson in proton\textendash{}proton collisions at
  $\sqrt{s}= $ 13 $\,\text {TeV}$}},
  \href{https://doi.org/10.1140/epjc/s10052-018-6049-9}{\emph{Eur. Phys. J. C}
  {\bfseries 78} (2018) 589}
  [\href{https://arxiv.org/abs/1712.09814}{{\ttfamily 1712.09814}}].

\bibitem{ATLAS:2020nzk}
{\scshape ATLAS} collaboration, \emph{{Differential cross-section measurements
  for the electroweak production of dijets in association with a $Z$ boson in
  proton\textendash{}proton collisions at ATLAS}},
  \href{https://doi.org/10.1140/epjc/s10052-020-08734-w}{\emph{Eur. Phys. J. C}
  {\bfseries 81} (2021) 163}
  [\href{https://arxiv.org/abs/2006.15458}{{\ttfamily 2006.15458}}].

\bibitem{Bellan:2021dcy}
R.~Bellan et~al., \emph{{A sensitivity study of VBS and diboson WW to
  dimension-6 EFT operators at the LHC}},
  \href{https://doi.org/10.1007/JHEP05(2022)039}{\emph{JHEP} {\bfseries 05}
  (2022) 039} [\href{https://arxiv.org/abs/2108.03199}{{\ttfamily
  2108.03199}}].

\bibitem{Eboli:2006wa}
O.J.P.~Eboli, M.C.~Gonzalez-Garcia and J.K.~Mizukoshi, \emph{{p p $\rightarrow$
  j j e$^\pm$ $\mu^\pm \nu \nu$ and j j e$^\pm$ $\mu^\pm \nu \nu$ at
  O($\alpha_{em}^6$) and O($\alpha_{em}^4\alpha_{s}^2$) for the study of the
  quartic electroweak gauge boson vertex at CERN LHC}},
  \href{https://doi.org/10.1103/PhysRevD.74.073005}{\emph{Phys. Rev. D}
  {\bfseries 74} (2006) 073005}
  [\href{https://arxiv.org/abs/hep-ph/0606118}{{\ttfamily hep-ph/0606118}}].

\bibitem{Almeida:2020ylr}
E.d.S.~Almeida, O.J.P.~\'Eboli and M.C.~Gonzalez\textendash{}Garcia,
  \emph{{Unitarity constraints on anomalous quartic couplings}},
  \href{https://doi.org/10.1103/PhysRevD.101.113003}{\emph{Phys. Rev. D}
  {\bfseries 101} (2020) 113003}
  [\href{https://arxiv.org/abs/2004.05174}{{\ttfamily 2004.05174}}].

\bibitem{Feruglio:1992wf}
F.~Feruglio, \emph{{The Chiral approach to the electroweak interactions}},
  \href{https://doi.org/10.1142/S0217751X93001946}{\emph{Int. J. Mod. Phys. A}
  {\bfseries 8} (1993) 4937}
  [\href{https://arxiv.org/abs/hep-ph/9301281}{{\ttfamily hep-ph/9301281}}].

\bibitem{Grinstein:2007iv}
B.~Grinstein and M.~Trott, \emph{{A Higgs-Higgs bound state due to new physics
  at a TeV}}, \href{https://doi.org/10.1103/PhysRevD.76.073002}{\emph{Phys.
  Rev. D} {\bfseries 76} (2007) 073002}
  [\href{https://arxiv.org/abs/0704.1505}{{\ttfamily 0704.1505}}].

\bibitem{Brivio:2013pma}
I.~Brivio, T.~Corbett, O.J.P.~\'Eboli, M.B.~Gavela, J.~Gonzalez-Fraile,
  M.C.~Gonzalez-Garcia et~al., \emph{{Disentangling a dynamical Higgs}},
  \href{https://doi.org/10.1007/JHEP03(2014)024}{\emph{JHEP} {\bfseries 03}
  (2014) 024} [\href{https://arxiv.org/abs/1311.1823}{{\ttfamily 1311.1823}}].

\bibitem{Brivio:2016fzo}
I.~Brivio, J.~Gonzalez-Fraile, M.C.~Gonzalez-Garcia and L.~Merlo, \emph{{The
  complete HEFT Lagrangian after the LHC Run I}},
  \href{https://doi.org/10.1140/epjc/s10052-016-4211-9}{\emph{Eur. Phys. J. C}
  {\bfseries 76} (2016) 416}
  [\href{https://arxiv.org/abs/1604.06801}{{\ttfamily 1604.06801}}].

\bibitem{CMS:2020gfh}
{\scshape CMS} collaboration, \emph{{Measurements of production cross sections
  of WZ and same-sign WW boson pairs in association with two jets in
  proton-proton collisions at $\sqrt{s} =$ 13 TeV}},
  \href{https://doi.org/10.1016/j.physletb.2020.135710}{\emph{Phys. Lett. B}
  {\bfseries 809} (2020) 135710}
  [\href{https://arxiv.org/abs/2005.01173}{{\ttfamily 2005.01173}}].

\bibitem{CMS:2019qfk}
{\scshape CMS} collaboration, \emph{{Search for anomalous electroweak
  production of vector boson pairs in association with two jets in
  proton-proton collisions at 13 TeV}},
  \href{https://doi.org/10.1016/j.physletb.2019.134985}{\emph{Phys. Lett. B}
  {\bfseries 798} (2019) 134985}
  [\href{https://arxiv.org/abs/1905.07445}{{\ttfamily 1905.07445}}].

\bibitem{CMS:2020ioi}
{\scshape CMS} collaboration, \emph{{Measurement of the cross section for
  electroweak production of a Z boson, a photon and two jets in proton-proton
  collisions at $\sqrt{s} =$ 13 TeV and constraints on anomalous quartic
  couplings}}, \href{https://doi.org/10.1007/JHEP06(2020)076}{\emph{JHEP}
  {\bfseries 06} (2020) 076}
  [\href{https://arxiv.org/abs/2002.09902}{{\ttfamily 2002.09902}}].

\bibitem{CMS:2020ypo}
{\scshape CMS} collaboration, \emph{{Observation of electroweak production of
  W$\gamma$ with two jets in proton-proton collisions at $\sqrt {s}$ = 13
  TeV}}, \href{https://doi.org/10.1016/j.physletb.2020.135988}{\emph{Phys.
  Lett. B} {\bfseries 811} (2020) 135988}
  [\href{https://arxiv.org/abs/2008.10521}{{\ttfamily 2008.10521}}].

\bibitem{Baglio:2015eon}
J.~Baglio, \emph{{Next-To-Leading Order QCD Corrections to Associated
  Production of a SM Higgs Boson with a Pair of Weak Bosons in the
  POWHEG-BOX}}, \href{https://doi.org/10.1103/PhysRevD.93.054010}{\emph{Phys.
  Rev. D} {\bfseries 93} (2016) 054010}
  [\href{https://arxiv.org/abs/1512.05787}{{\ttfamily 1512.05787}}].

\bibitem{Baglio:2016ofi}
J.~Baglio, \emph{{Gluon fusion and $b\bar{b}$ corrections to $H W^+ W^- / H Z
  Z$ production in the POWHEG-BOX}},
  \href{https://doi.org/10.1016/j.physletb.2016.10.066}{\emph{Phys. Lett. B}
  {\bfseries 764} (2017) 54}
  [\href{https://arxiv.org/abs/1609.05907}{{\ttfamily 1609.05907}}].

\bibitem{Brass:2018hfw}
S.~Brass, C.~Fleper, W.~Kilian, J.~Reuter and M.~Sekulla, \emph{{Transversal
  Modes and Higgs Bosons in Electroweak Vector-Boson Scattering at the LHC}},
  \href{https://doi.org/10.1140/epjc/s10052-018-6398-4}{\emph{Eur. Phys. J. C}
  {\bfseries 78} (2018) 931}
  [\href{https://arxiv.org/abs/1807.02512}{{\ttfamily 1807.02512}}].

\bibitem{Rauch:2016pai}
M.~Rauch, \emph{{Vector-Boson Fusion and Vector-Boson Scattering}},
  \href{https://arxiv.org/abs/1610.08420}{{\ttfamily 1610.08420}}.

\bibitem{Covarelli:2021gyz}
R.~Covarelli, M.~Pellen and M.~Zaro, \emph{{Vector-Boson scattering at the LHC:
  Unraveling the electroweak sector}},
  \href{https://doi.org/10.1142/S0217751X2130009X}{\emph{Int. J. Mod. Phys. A}
  {\bfseries 36} (2021) 2130009}
  [\href{https://arxiv.org/abs/2102.10991}{{\ttfamily 2102.10991}}].

\bibitem{aQGCcomp}
CMS\hspace{1mm}Collaboration, ``{Limits on anomalous triple and quartic gauge
  couplings}.''
  \url{https://twiki.cern.ch/twiki/bin/view/CMSPublic/PhysicsResultsSMPaTGC}.

\bibitem{Perez:2018kav}
G.~Perez, M.~Sekulla and D.~Zeppenfeld, \emph{{Anomalous quartic gauge
  couplings and unitarization for the vector boson scattering process
  $pp\rightarrow W^+W^+jjX\rightarrow \ell ^+\nu _\ell \ell ^+\nu _\ell jjX$}},
  \href{https://doi.org/10.1140/epjc/s10052-018-6230-1}{\emph{Eur. Phys. J. C}
  {\bfseries 78} (2018) 759}
  [\href{https://arxiv.org/abs/1807.02707}{{\ttfamily 1807.02707}}].

\bibitem{CMS:2021ssj}
{\scshape CMS} collaboration, \emph{{Search for Higgs boson pair production in
  the four b quark final state in proton-proton collisions at $\sqrt{s}$ = 13
  TeV}},  \href{https://arxiv.org/abs/2202.09617}{{\ttfamily 2202.09617}}.

\bibitem{CMS:2020tkr}
{\scshape CMS} collaboration, \emph{{Search for nonresonant Higgs boson pair
  production in final states with two bottom quarks and two photons in
  proton-proton collisions at $ \sqrt{s} $ = 13 TeV}},
  \href{https://doi.org/10.1007/JHEP03(2021)257}{\emph{JHEP} {\bfseries 03}
  (2021) 257} [\href{https://arxiv.org/abs/2011.12373}{{\ttfamily
  2011.12373}}].

\bibitem{CMS:2022pga}
{\scshape CMS} collaboration, ``{Search for nonresonant Higgs boson pair
  production in final states with two bottom quarks and two tau leptons in
  proton-proton collisions at $\sqrt{s} = 13~\mathrm{TeV}$,
  CMS-PAS-HIG-20-010}.'' 2022.

\bibitem{ATLAS:2020jgy}
{\scshape ATLAS} collaboration, \emph{{Search for the $HH \rightarrow b \bar{b}
  b \bar{b}$ process via vector-boson fusion production using proton-proton
  collisions at $\sqrt{s} = 13$ TeV with the ATLAS detector}},
  \href{https://doi.org/10.1007/JHEP07(2020)108}{\emph{JHEP} {\bfseries 07}
  (2020) 108} [\href{https://arxiv.org/abs/2001.05178}{{\ttfamily
  2001.05178}}].

\bibitem{Bishara:2016kjn}
F.~Bishara, R.~Contino and J.~Rojo, \emph{{Higgs pair production in
  vector-boson fusion at the LHC and beyond}},
  \href{https://doi.org/10.1140/epjc/s10052-017-5037-9}{\emph{Eur. Phys. J. C}
  {\bfseries 77} (2017) 481}
  [\href{https://arxiv.org/abs/1611.03860}{{\ttfamily 1611.03860}}].

\bibitem{CMS:2022dwd}
{\scshape CMS} collaboration, \emph{{A portrait of the Higgs boson by the CMS
  experiment ten years after the discovery}},
  \href{https://doi.org/10.1038/s41586-022-04892-x}{\emph{Nature} {\bfseries
  607} (2022) 60} [\href{https://arxiv.org/abs/2207.00043}{{\ttfamily
  2207.00043}}].

\bibitem{Contino:2016jqw}
R.~Contino, A.~Falkowski, F.~Goertz, C.~Grojean and F.~Riva, \emph{{On the
  Validity of the Effective Field Theory Approach to SM Precision Tests}},
  \href{https://doi.org/10.1007/JHEP07(2016)144}{\emph{JHEP} {\bfseries 07}
  (2016) 144} [\href{https://arxiv.org/abs/1604.06444}{{\ttfamily
  1604.06444}}].

\bibitem{Garcia-Garcia:2019oig}
C.~Garcia-Garcia, M.~Herrero and R.A.~Morales, \emph{{Unitarization effects in
  EFT predictions of WZ scattering at the LHC}},
  \href{https://doi.org/10.1103/PhysRevD.100.096003}{\emph{Phys. Rev. D}
  {\bfseries 100} (2019) 096003}
  [\href{https://arxiv.org/abs/1907.06668}{{\ttfamily 1907.06668}}].

\bibitem{Kalinowski:2018oxd}
J.~Kalinowski, P.~Koz\'ow, S.~Pokorski, J.~Rosiek, M.~Szleper and S.~Tkaczyk,
  \emph{{Same-sign WW scattering at the LHC: can we discover BSM effects before
  discovering new states?}},
  \href{https://doi.org/10.1140/epjc/s10052-018-5885-y}{\emph{Eur. Phys. J. C}
  {\bfseries 78} (2018) 403}
  [\href{https://arxiv.org/abs/1802.02366}{{\ttfamily 1802.02366}}].

\bibitem{Kozow:2019txg}
P.~Koz\'ow, L.~Merlo, S.~Pokorski and M.~Szleper, \emph{{Same-sign WW
  Scattering in the HEFT: Discoverability vs. EFT Validity}},
  \href{https://doi.org/10.1007/JHEP07(2019)021}{\emph{JHEP} {\bfseries 07}
  (2019) 021} [\href{https://arxiv.org/abs/1905.03354}{{\ttfamily
  1905.03354}}].

\bibitem{Chaudhary:2019aim}
G.~Chaudhary, J.~Kalinowski, M.~Kaur, P.~Koz\'ow, K.~Sandeep, M.~Szleper
  et~al., \emph{{EFT triangles in the same-sign $WW$ scattering process at the
  HL-LHC and HE-LHC}},
  \href{https://doi.org/10.1140/epjc/s10052-020-7728-x}{\emph{Eur. Phys. J. C}
  {\bfseries 80} (2020) 181}
  [\href{https://arxiv.org/abs/1906.10769}{{\ttfamily 1906.10769}}].

\bibitem{Alwall:2014hca}
J.~Alwall, R.~Frederix, S.~Frixione, V.~Hirschi, F.~Maltoni, O.~Mattelaer
  et~al., \emph{{The automated computation of tree-level and next-to-leading
  order differential cross sections, and their matching to parton shower
  simulations}}, \href{https://doi.org/10.1007/JHEP07(2014)079}{\emph{JHEP}
  {\bfseries 07} (2014) 079} [\href{https://arxiv.org/abs/1405.0301}{{\ttfamily
  1405.0301}}].

\bibitem{Frederix:2018nkq}
R.~Frederix, S.~Frixione, V.~Hirschi, D.~Pagani, H.S.~Shao and M.~Zaro,
  \emph{{The automation of next-to-leading order electroweak calculations}},
  \href{https://doi.org/10.1007/JHEP11(2021)085}{\emph{JHEP} {\bfseries 07}
  (2018) 185} [\href{https://arxiv.org/abs/1804.10017}{{\ttfamily
  1804.10017}}].

\bibitem{Frederix:2011qg}
R.~Frederix, S.~Frixione, V.~Hirschi, F.~Maltoni, R.~Pittau and P.~Torrielli,
  \emph{{W and $Z/\gamma*$ boson production in association with a
  bottom-antibottom pair}},
  \href{https://doi.org/10.1007/JHEP09(2011)061}{\emph{JHEP} {\bfseries 09}
  (2011) 061} [\href{https://arxiv.org/abs/1106.6019}{{\ttfamily 1106.6019}}].

\bibitem{Frederix:2014hta}
R.~Frederix, S.~Frixione, V.~Hirschi, F.~Maltoni, O.~Mattelaer, P.~Torrielli
  et~al., \emph{{Higgs pair production at the LHC with NLO and parton-shower
  effects}}, \href{https://doi.org/10.1016/j.physletb.2014.03.026}{\emph{Phys.
  Lett. B} {\bfseries 732} (2014) 142}
  [\href{https://arxiv.org/abs/1401.7340}{{\ttfamily 1401.7340}}].

\bibitem{Bagnaschi:2018dnh}
E.~Bagnaschi, F.~Maltoni, A.~Vicini and M.~Zaro, \emph{{Lepton-pair production
  in association with a $ b\overline{b} $ pair and the determination of the $W$
  boson mass}}, \href{https://doi.org/10.1007/JHEP07(2018)101}{\emph{JHEP}
  {\bfseries 07} (2018) 101}
  [\href{https://arxiv.org/abs/1803.04336}{{\ttfamily 1803.04336}}].

\bibitem{Ballestrero:2018anz}
A.~Ballestrero et~al., \emph{{Precise predictions for same-sign W-boson
  scattering at the LHC}},
  \href{https://doi.org/10.1140/epjc/s10052-018-6136-y}{\emph{Eur. Phys. J. C}
  {\bfseries 78} (2018) 671}
  [\href{https://arxiv.org/abs/1803.07943}{{\ttfamily 1803.07943}}].

\bibitem{Ball:2013hta}
{\scshape NNPDF} collaboration, \emph{{Parton distributions with QED
  corrections}},
  \href{https://doi.org/10.1016/j.nuclphysb.2013.10.010}{\emph{Nucl. Phys. B}
  {\bfseries 877} (2013) 290}
  [\href{https://arxiv.org/abs/1308.0598}{{\ttfamily 1308.0598}}].

\bibitem{Degrande:2011ua}
C.~Degrande, C.~Duhr, B.~Fuks, D.~Grellscheid, O.~Mattelaer and T.~Reiter,
  \emph{{UFO - The Universal FeynRules Output}},
  \href{https://doi.org/10.1016/j.cpc.2012.01.022}{\emph{Comput. Phys. Commun.}
  {\bfseries 183} (2012) 1201}
  [\href{https://arxiv.org/abs/1108.2040}{{\ttfamily 1108.2040}}].

\bibitem{twikiEboli}
\url{https://feynrules.irmp.ucl.ac.be/wiki/AnomalousGaugeCoupling}.

\bibitem{Arnold:2008rz}
K.~Arnold et~al., \emph{{VBFNLO: A Parton level Monte Carlo for processes with
  electroweak bosons}},
  \href{https://doi.org/10.1016/j.cpc.2009.03.006}{\emph{Comput. Phys. Commun.}
  {\bfseries 180} (2009) 1661}
  [\href{https://arxiv.org/abs/0811.4559}{{\ttfamily 0811.4559}}].

\bibitem{Bierlich:2022pfr}
C.~Bierlich et~al., \emph{{A comprehensive guide to the physics and usage of
  PYTHIA 8.3}},  \href{https://arxiv.org/abs/2203.11601}{{\ttfamily
  2203.11601}}.

\bibitem{CMS:2021ugl}
{\scshape CMS} collaboration, \emph{{Measurements of production cross sections
  of the Higgs boson in the four-lepton final state in
  proton\textendash{}proton collisions at $\sqrt{s} = 13\,\text {Te}\text {V}
  $}}, \href{https://doi.org/10.1140/epjc/s10052-021-09200-x}{\emph{Eur. Phys.
  J. C} {\bfseries 81} (2021) 488}
  [\href{https://arxiv.org/abs/2103.04956}{{\ttfamily 2103.04956}}].

\bibitem{Feldman:1997qc}
G.J.~Feldman and R.D.~Cousins, \emph{{A Unified approach to the classical
  statistical analysis of small signals}},
  \href{https://doi.org/10.1103/PhysRevD.57.3873}{\emph{Phys. Rev. D}
  {\bfseries 57} (1998) 3873}
  [\href{https://arxiv.org/abs/physics/9711021}{{\ttfamily physics/9711021}}].

\end{thebibliography}\endgroup

\end{document}